\documentclass[article,a4paper,10pt]{article}

\usepackage[affil-it]{authblk}
\usepackage{amsmath}
\usepackage{times}
\usepackage{bm}
\usepackage{bbm}
\usepackage{natbib}
\usepackage{tikz}
\usepackage{algorithm}
\usepackage{setspace}
\usepackage{fullpage}
\usepackage{multirow}
\usepackage{framed}

\newcommand{\bfx}{\boldsymbol{x}}
\newcommand{\bfm}{\boldsymbol{m}}
\newcommand{\bfe}{\boldsymbol{e}}

\newcommand{\bfz}{\boldsymbol{z}}
\newcommand{\bfy}{\boldsymbol{y}}
\newcommand{\bfs}{\boldsymbol{s}}
\newcommand{\bfpi}{\boldsymbol{\pi}}
\newcommand{\bfphi}{\boldsymbol{\phi}}
\newcommand{\bfmu}{\boldsymbol{\mu}}
\newcommand{\bftheta}{\boldsymbol{\theta}}

\begin{document}

\title{Statistical Inference in Hidden Markov Models using $k$-segment Constraints}

\author[1]{Michalis K. Titsias\thanks{E-mail: mtitsias@aueb.gr}}
\author[2]{Christopher Yau\thanks{E-mail: cyau@well.ox.ac.uk}}
\author[3]{Christopher C. Holmes\thanks{E-mail: cholmes@stats.ox.ac.uk}}
\affil[1]{Athens University of Economics and Business, 76, Patission Str. GR10434, Athens, Greece}
\affil[2]{Wellcome Trust Centre for Human Genetics, University of Oxford, Roosevelt Drive, Oxford, United Kingdom}
\affil[3]{Department of Statistics, University of Oxford, 1 South Parks Road, Oxford, United Kingdom}

\maketitle

\begin{abstract}
Hidden Markov models (HMMs) are one of the most widely used statistical methods for analyzing sequence data. However, the reporting of output from HMMs has largely been restricted to the presentation of the most-probable (MAP) hidden state sequence, found via the Viterbi algorithm, or the sequence of most probable marginals using the forward-backward (F-B) algorithm. In this article, we expand the amount of information we could obtain from the posterior distribution of an HMM by introducing linear-time dynamic programming algorithms that, we collectively call $k$-segment algorithms, that allow us to i) find MAP sequences, ii) compute posterior probabilities and iii) simulate sample paths conditional on a user specified number of segments, i.e. contiguous runs in a hidden state, possibly of a particular type. We illustrate the utility of these methods using simulated and real examples and highlight the application of prospective and retrospective use of these methods for fitting HMMs or exploring existing model fits.
\end{abstract}

\section{Introduction}

The use of the Hidden Markov Model (HMM) is ubiquitous in a range of sequence analysis applications across a range of scientific and engineering domains, including signal processing \citep{juang1991hidden,crouse1998wavelet}, genomics \citep{eddy1998profile,li2003modeling} and finance \citep{paas2007discrete}. Fundamentally, the HMM is a mixture model whose mixing distribution is a finite state Markov chain \citep{Rabiner89, Cappe2005}. Whilst the Markov assumptions rarely correspond to the true physical generative process, it often adequately captures first-order properties that make it a useful approximating model for sequence data in many instances whilst remaining tractable even for very large datasets. As a consequence, HMM-based algorithms can give highly competitive performance in many applications.

Central to the tractability of HMMs is the availability of recursive algorithms that allow fundamental quantities to be computed efficiently \citep{baum66,Viterbi67}. These include the Viterbi algorithm which computes the most probable hidden state sequence and the forward-backward algorithm which computes the marginal probability of a given state at a point in the sequence. Computation for the HMM has been well-summarized in the comprehensive and widely read tutorial by \cite{Rabiner89} with a Bayesian treatment given more recently by \cite{Scott2002}. It is a testament to the completeness of these recursive methods that there have been few generic additions to the HMM toolbox since these were first described in the 1960s. However, as HMM approaches continue to be applied in increasingly diverse scientific domains and ever larger data sets, there is interest in expanding the generic toolbox available for HMM inference to encompass unmet needs. 

The motivation for our work is to develop mechanisms to allow the \emph{exploration} of the posterior sequence space. Typically, standard HMM inference limits itself to reporting a few standard quantities. For an $M$-state Markov chain of length $N$ there exists of $M^N$ possible sequences but often only the most probable sequence or the $N M$ marginal posterior probabilities are used to summarize the whole posterior distribution. Yet, it is clear that, when the state space is large and/or the sequences long, many other sequences maybe of interest. Modifications of the Viterbi algorithm can allow arbitrary number of the most probable sequences to be enumerated whilst Bayesian techniques allows us to sample sequences from the posterior distribution. However, since a small change to the most likely sequences typically give new sequences with similar probability, these approaches do not lead to reports of \emph{qualitatively diverse} sequences. By which we mean, alternative sequence predictions that might lead to different decisions or scientific conclusions.

In this article we describe a set of novel recursive methods for HMM computation that incorporates segmental constraints that we call \emph{$k$-segment inference algorithms}. These are so-called because the algorithms are constrained to consider only sequences involving no more than $k-1$ specified transition events. We show that $k$-segment procedures provide an intuitive approach for posterior exploration of the sequence space allowing diverse sequence predictions containing $1, 2, \dots, $ and $k$ segments or specific transitions of interest. These methods can be applied prospectively during model fitting or retrospectively to an existing model. In the latter case, the utility of the methods described here comes at no cost (other than computational time) to the HMM user and we provide illustrative examples to highlight novel insights that maybe gained through $k$-segment approaches.

\section{Background}

The HMM encodes for two types of random sequences: the hidden state sequence or path $\bfx = (x_1,\ldots,x_N)$ and the observed data sequence $\bfy = (y_1, \ldots, y_N)$. Individual hidden states take discrete values, such that  $x_n \in \{1,\ldots,M\}$, while observed variables can be of arbitrary type. The hidden state sequence  $\bfx$ follows a Markov chain so that   
\begin{equation}
	p(\bfx|\bfpi_0, A) = p(x_1|\bfpi_0) \prod_{n=2}^N p(x_n|x_{n-1},A).
\label{eq:markovchain}  
\end{equation}
Here, the first hidden state $x_1$ is drawn from some initial probability vector $\bfpi_0$ so that $\pi_{0,m} = p(x_1=m)$ denotes the probability of $x_1$ being in state $m \in \{1,\ldots,M\}$, whereas any subsequent hidden state $x_n$ (with $n > 1$) is drawn according to a transition matrix $A$ so that $[A]_{m' m} = p(x_n=m|x_{n-1}=m')$ expresses the probability of moving to a state $m$ from $m'$. Given a path $\bfx$ following the Markov chain in (\ref{eq:markovchain}), the observed data are generated independently according to 
\begin{equation}
	p(\bfy |\bfx) = \prod_{n=1}^N p(y_n|x_n, \bfphi),
	\label{eq:likelihood}
\end{equation}  
where the densities $p(y_n|x_n=m,\bfphi), m=1,\ldots,M$, are often referred to as the emission densities and are parametrized by $\bfphi$. Next we shall collectively denote all HMM parameters, i.e. $\bfpi_0$, $A$ and $\bfphi$, by $\bftheta$.

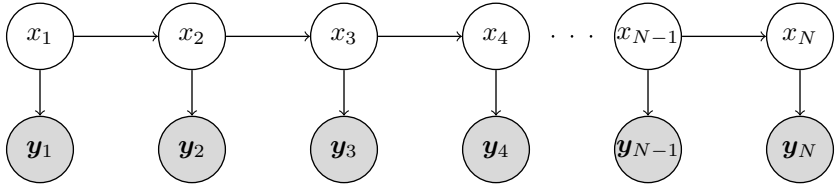
\begin{figure}[!t]
	\begin{center}
	\begin{tikzpicture}
	\tikzstyle{every node} = [shape=circle, inner sep=0pt, line width=0.5, font=\fontsize{10}{10},  minimum size=25pt, draw]
	\node (b1) at (0,-1.5)  {$x_1$};
	\node (b2) at (2,-1.5) {$x_2$};
	\node (b3) at (4,-1.5) {$x_3$};
	\node (b4) at (6,-1.5) {$x_4$};
	\node (b5) at (8,-1.5) {$x_{N-1}$};
	\node (b6) at (10,-1.5) {$x_N$}; 
	\node[fill=gray!30] (y1) at (0,-3)  {$\bfy_1$};
	\node[fill=gray!30] (y2) at (2,-3) {$\bfy_2$};
	\node[fill=gray!30] (y3) at (4,-3) {$\bfy_3$};
	\node[fill=gray!30] (y4) at (6,-3) {$\bfy_4$};
	\node[fill=gray!30] (y5) at (8,-3) {$\bfy_{N-1}$};
	\node[fill=gray!30] (y6) at (10,-3) {$\bfy_N$}; 
	\draw[line width=0.5] [->] (b1) -- (b2);
	\draw[line width=0.5] [->] (b2) -- (b3); 
	\draw[line width=0.5] [->] (b3) -- (b4);
	\draw[line width=0.5] [->] (b5) -- (b6);
	\draw[line width=0.5] [->] (b1) -- (y1);
	\draw[line width=0.5] [->] (b2) -- (y2); 
	\draw[line width=0.5] [->] (b3) -- (y3);
	\draw[line width=0.5] [->] (b4) -- (y4);
	\draw[line width=0.5] [->] (b5) -- (y5);
	\draw[line width=0.5] [->] (b6) -- (y6);
	\draw[fill=black] (6.75,-1.5) circle (0.1mm);
	\draw[fill=black] (7,-1.5) circle (0.1mm); 
	\draw[fill=black] (7.25,-1.5) circle (0.1mm);
	\end{tikzpicture}
	\end{center}
	\caption{HMM depicted as a directed graphical model. \label{fig:HMM}}
\end{figure}

Statistical estimation in HMMs takes advantage of the Markov dependence structure, shown in Figure \ref{fig:HMM}, which allows efficient dynamic programming algorithms to be applied. For instance,   maximum likelihood (ML) over the parameters $\bftheta$ via the EM algorithm is carried out by the forward-backward (F-B) recursion \citep{baum66} that implements the Expectation step in $O(M^2 N)$ time. A similar recursion having the same time complexity is the Viterbi algorithm \citep{Viterbi67} which, given a fixed value for the parameters, estimates the maximum {\it a posteriori} (MAP) hidden sequence. Furthermore,  straightforward generalizations of the Viterbi algorithm estimate the $P$-best list of most probable sequences \citep{Schwartz_Chow90, Nilsson_and_Goldberger:2001}. In contrast to ML point estimation, a Bayesian approach assigns a prior distribution $p(\bftheta)$ over the parameters and seeks to estimate expectations taken under the posterior distribution $p(\bfx,\bftheta|\bfy)$. The Bayesian framework also greatly benefits from efficient recursions derived as  subroutines of Monte Carlo algorithms.  Specifically, the popular Gibbs sampling scheme \citep{Scott2002} relies on the forward-filtering-backward-sampling (FF-BS) recursion that simulates in $O(M^2 N)$ time  a hidden sequence from the conditional posterior distribution $p(\bfx|\bftheta,\bfy)$. In summary, all recursions  mentioned above have linear time complexity with respect to the length of the sequence $N$  and are instances of more general inference tools developed in the theory of probabilistic graphical models \citep{cowell2002prob_networks,koller-Friedman09}.

\section{Motivation}

While the linear time efficiency of the current HMM recursions is one of the keys for the widespread adoption of HMMs in applications, the information regarding the posterior distribution obtained by these algorithms is still very limited. To this end, we define novel probabilistic inference problems for exploration of the HMM posterior distribution and we efficiently solve these problems by introducing linear time recursions. To start with a motivating example, assume that we are interested in the event
\begin{equation}
	c_{\bfx} = \#\{\bfx \ \text{contains transitions of a certain class} \},
	\label{eq:sxtrans1}
\end{equation}
which denotes the number of times a certain class of transitions occurs along the hidden path $\bfx$ of the
HMM. Then, we may wish to compute the probability: 
\begin{equation}
	p(c_{\bfx}=k|\bfy,\bftheta) = \sum_{\bfx}  I( c_{\bfx} =k ) p(\bfx|\bftheta,\bfy), 
	\label{eq:sxequalk}
\end{equation}
which is a global marginal obtained after a summation over all paths having  exactly $k$ occurrences from the certain class of transitions. This probability cannot be obtained from the current F-B recursion which allows us to compute only local marginals such as $p(x_n|\bfy,\bftheta)$ or $p(x_{n-1},x_n|\bfy,\bftheta)$. The use of Monte Carlo methods to approximate the right hand side of (\ref{eq:sxequalk}) is also unsuitable because, while fast and exact simulation from $p(\bfx|\bftheta,\bfy)$ is possible by means of the FF-BS recursion, the obtained accuracy could be insufficient when the underlying value of $p(c_{\bfx} =k|\bfy,\bftheta)$ is very small due to the extremeness of the event $c_{\bfx}=k$. 

We can also define several other related tasks that throughout the paper we collectively refer to as {\em $k$-segment inference problems}.  In such problems,  we  insert hard constraints into the HMM involving the number and type of segments we want to see along the path $\bfx$, and then we query the model to provide us with  probabilities or representative paths characterizing that constraint. Some additional representative examples of $k$-segment inference that we will study in this article are the computation of the optimal MAP sequence associated with the event $c_{\bfx} = k$ and the simulation of paths from the conditional distribution $p(\bfx| c_{\bfx} = k, \bfy,\bftheta)$. 

To solve  $k$-segment inference problems we develop efficient linear time dynamic programming recursions. The solution we provide introduces auxiliary counting variables into the original HMM so that we obtain an extended state-space HMM which is consistent with the original model. The auxiliary counting variables used in this augmentation allows us to inject evidence or constraints into the model so that the standard recursions applied to the extended HMM allow us to solve all $k$-segment inference problems of interest. This provides a simple and elegant solution and results in new HMM recursions that generalize the standard F-B, Viterbi and FF-BS algorithms.

The remaining of the article has as follows. Section \ref{sec:theory} describes the main theory of $k$-segment inference, for applications to {\it a posteriori} model exploration, and derives the novel HMM recursions.  Section \ref{sec:statinference} considers model fitting  under $k$-segment constraints and presents suitable EM and Bayesian learning procedures. Section \ref{sec:extensions} presents extensions to the basic $k$-segment problems, while Section \ref{sec:relatedwork} discusses related work. Section \ref{sec:experiments} considers sequence analysis using two real-world examples from cancer genomics and text information retrieval. Finally, Section \ref{sec:discussion} concludes with a discussion and directions for future work.

\section{Theory of $k$-segment inference \label{sec:theory}}

This section presents the theoretical foundations of $k$-segment inference problems starting with section \ref{sec:kseg} that defines such problems. Section \ref{sec:counting} reformulates these problems in terms of an extended state-space HMM having auxiliary counting variables and section \ref{sec:dynprog} presents efficient solutions  based on linear time recursions. 
Finally, Section \ref{sec:graphillustr} gives a graphical illustration of the proposed algorithms using a simulated sequence. 
All the algorithms described in this section assume a fixed setting for the parameters $\bftheta$. Therefore, to keep our expressions uncluttered in the following we drop $\bftheta$ from our expressions and write for instance $p(\bfx|\bfy,\bftheta)$ as $p(\bfx|\bfy)$ and $p(\bfy|\bftheta)$ as $p(\bfy)$. 

\subsection{$k$-segment inference problems \label{sec:kseg}}

Any hidden path $\bfx$ in a HMM can have from $0$ up to $N-1$ transitions or equivalently from  $1$ up to $N$ segments, where a segment is defined as a contiguous run of indices where $x_{n-1} = x_n$. Following the notation used in eq.\ (\ref{eq:sxtrans1}), we  define the number of all segments in $\bfx$ by
\begin{equation}
	c_{\bfx} = 1 + \sum_{n=2}^N I(x_{n-1} \neq x_n),
	\label{eq:sxtrans}
\end{equation}
where $I(\cdot)$ denotes the indicator function. $c_{\bfx}$ is the sum of the number of transitions, i.e.\ the locations in the hidden path where $x_{n-1} \neq x_n$, and the value one that accounts for the initial segment which is not the result of a transition. 

Subsets of hidden paths associated with different number of segments comprise exclusive events which allow to decompose the posterior  distribution $p(\bfx|\bfy)$ as follows. If we introduce the events $c_{\bfx} = k$,  with $k=1,\ldots,N$, each corresponding to the subset of paths $\{\bfx| c_{\bfx} = k \}$ having exactly $k$ segments, the posterior distribution $p(\bfx|\bfy)$ can be written as the following mixture: 
\begin{equation}
	p(\bfx|\bfy) = \sum_{k=1}^N 
	p(\bfx, c_{\bfx} = k|\bfy) = \sum_{k=1}^{N} p(\bfx | c_{\bfx} = k, \bfy) p(c_{\bfx} = k|\bfy),
	\label{eq:pbfxbfydecomp}
\end{equation} 
where
\begin{equation}
	p(\bfx|\bfy, c_{\bfx} = k) = \frac{I(c_{\bfx} = k) p(\bfy|\bfx) p(\bfx)} {\sum_{\bfx: c_{\bfx} = k } p(\bfy|\bfx ) p(\bfx )},
	\label{eq:pxEKy}
\end{equation}
is the posterior distribution conditional on having $k$ segments, while
\begin{equation}
	p(c_{\bfx} = k|\bfy) = \frac{p(c_{\bfx}=k,\bfy)}{p(\bfy)} = \frac{\sum_{\bfx: c_{\bfx} = k } p(\bfy|\bfx) p(\bfx)} { \sum_{\bfx} p(\bfy|\bfx) p(\bfx)},
\end{equation}
is the posterior probability of the event $c_{\bfx} = k$. 

The mixture decomposition in eq.\  (\ref{eq:pbfxbfydecomp}) suggests that one way to explore the posterior distribution of the HMM is to compute quantities associated with the components of this mixture. This leads to the $k$-segment inference problems which can be divided into the following three types of problems: 

\begin{itemize}

	\item {\bf Optimal decoding:} Find the MAP hidden path that has $k$ segments, that is the path with the maximum value of $p(\bfx | c_{\bfx} = k, \bfy)$.

	\item {\bf Probability computation:} Find the posterior probability of having $k$ segments, i.e.\ $p(c_{\bfx} = k|\bfy)$.

	\item {\bf Path sampling:} Draw independent samples from $p(\bfx | c_{\bfx} = k, \bfy)$.

\end{itemize}

To this end, in Section \ref{sec:dynprog} we introduce efficient linear time algorithms to solve all the above tasks together with several additional related tasks associated with more general events of the form $k_1 \leq c_{\bfx} \leq k_2$, where  $1 \leq k_1 < k_2 \leq N$, such as finding the MAP of $p(\bfx | c_{\bfx} > k, \bfy)$, sampling from $p(\bfx | c_{\bfx} > k, \bfy)$ and etc. These algorithms are based on a reformulation of the above $k$-segment inference problems that uses an extended state-space HMM containing auxiliary counting variables. 

\subsection{Auxiliary counting Markov chains\label{sec:counting}}
 
The basis of our algorithm is the augmentation of the Markov chain in (\ref{eq:markovchain}) with auxiliary variables that count the number of segments. Specifically, $c_{\bfx}$ from (\ref{eq:sxtrans}) can be considered as a counter that  scans the path $\bfx$ and it increments by one any time it encounters a transition. We  can represent this counting process with a $N$-dimensional vector of auxiliary variables $\bfs$ which is an increasingly monotone sequence of non-negative integers. 

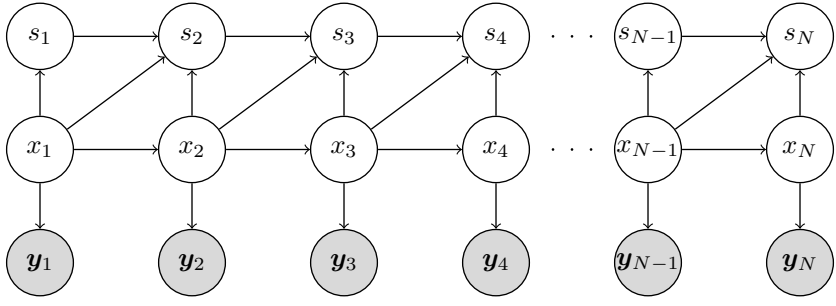
\begin{figure}
	\begin{center}
	\begin{tikzpicture}
	\tikzstyle{every node} = [shape=circle, inner sep=0pt, line width=0.5, font=\fontsize{10}{10},  minimum size=25pt, draw]
	\node (a1) at (0,0)  {$s_1$};
	\node (a2) at (2,0) {$s_2$};
	\node (a3) at (4,0) {$s_3$};
	\node (a4) at (6,0) {$s_4$};
	\node (a5) at (8,0) {$s_{N-1}$};
	\node (a6) at (10,0) {$s_N$}; 
	\node (b1) at (0,-1.5)  {$x_1$};
	\node (b2) at (2,-1.5) {$x_2$};
	\node (b3) at (4,-1.5) {$x_3$};
	\node (b4) at (6,-1.5) {$x_4$};
	\node (b5) at (8,-1.5) {$x_{N-1}$};
	\node (b6) at (10,-1.5) {$x_N$}; 
	\node[fill=gray!30] (y1) at (0,-3)  {$\bfy_1$};
	\node[fill=gray!30] (y2) at (2,-3) {$\bfy_2$};
	\node[fill=gray!30] (y3) at (4,-3) {$\bfy_3$};
	\node[fill=gray!30] (y4) at (6,-3) {$\bfy_4$};
	\node[fill=gray!30] (y5) at (8,-3) {$\bfy_{N-1}$};
	\node[fill=gray!30] (y6) at (10,-3) {$\bfy_N$}; 
	\draw[line width=0.5] [->] (a1) -- (a2);
	\draw[line width=0.5] [->] (a2) -- (a3);
	\draw[line width=0.5] [->] (a3) -- (a4);
	\draw[line width=0.5] [->] (a5) -- (a6);
	\draw[line width=0.5] [->] (b1) -- (a1);
	\draw[line width=0.5] [->] (b2) -- (a2);
	\draw[line width=0.5] [->] (b3) -- (a3);
	\draw[line width=0.5] [->] (b4) -- (a4);
	\draw[line width=0.5] [->] (b5) -- (a5);
	\draw[line width=0.5] [->] (b6) -- (a6);
	\draw[line width=0.5] [->] (b1) -- (b2);
	\draw[line width=0.5] [->] (b2) -- (b3); 
	\draw[line width=0.5] [->] (b3) -- (b4);
	\draw[line width=0.5] [->] (b5) -- (b6);
	\draw[line width=0.5] [->] (b1) -- (a2);
	\draw[line width=0.5] [->] (b2) -- (a3);
	\draw[line width=0.5] [->] (b3) -- (a4);
	\draw[line width=0.5] [->] (b5) -- (a6); 
	\draw[line width=0.5] [->] (b1) -- (y1);
	\draw[line width=0.5] [->] (b2) -- (y2); 
	\draw[line width=0.5] [->] (b3) -- (y3);
	\draw[line width=0.5] [->] (b4) -- (y4);
	\draw[line width=0.5] [->] (b5) -- (y5);
	\draw[line width=0.5] [->] (b6) -- (y6);
	\draw[fill=black] (6.75,0) circle (0.1mm);
	\draw[fill=black] (7,0) circle (0.1mm); 
	\draw[fill=black] (7.25,0) circle (0.1mm);
	\draw[fill=black] (6.75,-1.5) circle (0.1mm);
	\draw[fill=black] (7,-1.5) circle (0.1mm); 
	\draw[fill=black] (7.25,-1.5) circle (0.1mm);
	\end{tikzpicture}
	\end{center}
	\caption{Directed graphical model for the HMM augmented with 
	the counting chain. \label{fig:expandedHMM}}
\end{figure}

Conditioning on a certain path $\bfx$, $\bfs$ is sampled deterministically according to the Markov chain 
\begin{eqnarray}
	p(\bfs|\bfx) 	& = & p(s_1|x_1) \prod_{n=2}^N p(s_n|s_{n-1},x_{n-1},x_n), \nonumber \\
					& = & \delta_{s_1,1} \prod_{n=2}^N \left[ I(x_{n-1} \neq x_n) \delta_{s_n,s_{n-1}+1} + (1 - I(x_{n-1} \neq x_n) ) \delta_{s_n,s_{n-1}} \right],
	\label{eq:priors}
\end{eqnarray} 
where $\delta_{i,j}$ is the delta mass that equals one when $i=j$ and zero otherwise. We refer to the above conditional distribution as the {\em counting Markov chain} or counting chain because it is Markov chain that makes precise the concept of counting the segments. The counting chain starts at one, i.e.\ $s_1=1$ (which can be interpreted as sampling from the delta mass $\delta_{s_1,1}$), and then it increments by one so that $s_n = s_{n-1}+1$ every time a transition occurs in the hidden path, i.e.\ whenever $x_{n-1} \neq x_n$ which implies the generation of a new segment. The joint density of the HMM is augmented with the counting chain so that 
\begin{equation}
	p(\bfy,\bfx,\bfs) = p(\bfy|\bfx) p(\bfx) p(\bfs|\bfx),
	\label{eq:augmjoint}
\end{equation}
is the new joint density having the conditional independence structure shown as directed graphical model in Figure \ref{fig:expandedHMM}. Because this augmentation is consistent\footnote{Clearly, if we marginalize out $\bfs$ we recover correctly the joint density of the initial HMM.}, prior-to-posterior inference in the initial HMM and the HMM augmented with auxiliary variables are in theory  equivalent. However, in practice, inference in the latter model is more flexible since it 
allows to solve the $k$-segment inference problems through the insertion of constraints in the counting process. More precisely, the final value of the counter $s_N$ equals $c_{\bfx}$ so the event $c_{\bfx} = k$ can be realized by adding the evidence $s_N = k$ in the graphical model of Figure \ref{fig:expandedHMM}.  Therefore, all type of $k$-segment inference problems can be reformulated as follows: 

\begin{itemize}

	\item {\bf Optimal decoding:} The MAP hidden $\bfx^*$ of $p(\bfx | c_{\bfx} = k, \bfy)$ can be found according to 
	\begin{equation}
		(\bfx^*, \bfs_{\setminus N}^*) = \arg \max_{\bfx_*, \bfs_{\setminus N}} p(\bfy|\bfx)p(\bfx) p(\bfs_{\setminus N}, s_N=k|\bfx).
	\end{equation}
	
	\item {\bf Probability computation:} The posterior probability $p(c_{\bfx} = k|\bfy)$ can be expressed as $\frac{p(s_N=k, \bfy)}{p(\bfy)}$ where $p(\bfy)$ is known from the forward pass of the standard F-B algorithm and 
		\begin{equation}
			p(s_N=k, \bfy) = \sum_{\bfx,\bfs_{\setminus N}} p(\bfy|\bfx) p(\bfx) p(\bfs_{\setminus N}, s_N=k|\bfx).
			\label{eq:psNKy}
		\end{equation}

	\item {\bf Path sampling:} An independent sample $\widetilde{\bfx}$ from $p(\bfx | c_{\bfx} = k, \bfy)$ is obtained as 
		\begin{equation}
			(\widetilde{\bfx},\widetilde{\bfs}_{\setminus N})  \sim  p(\bfx,\bfs_{\setminus N}|s_N=k,\bfy)  \propto p(\bfy|\bfx) p(\bfx) p(\bfs_{\setminus N}, s_N=k|\bfx),
			\label{eq:samplekseg}
		\end{equation}
		
\end{itemize}
where in the above $\bfs_{\setminus N}$ denotes all counting variables apart from the final $s_N$ which is clamped to $k$. For more general events of the form $k_1 \leq s_N \leq k_2$, where $1 \leq k_1 < k_2 \leq N$, the above still holds with the slight modification that we will need additionally to maximize, marginalize or sample $s_N$, respectively for the three cases above, under the constraint $k_1 \leq s_N \leq k_2$. Simple proofs for the correctness of all above statements can be found in the Appendix \ref{app:proofs}.
  
Furthermore, the $k$-segment inference problems associated with the special case of the event $s_N > k$ can be equivalently reformulated by using a modified counting chain that absorbs when $s_n=k+1$, i.e.
\begin{equation}
	p(\bfs|\bfx) = \delta_{s_1,1} \prod_{n=2}^N \left[ I(x_n \neq x_{n-1} \ \& \ s_{n-1} \leq k) \delta_{s_n,s_{n-1}+1} + \left(1  -  I(x_n \neq x_{n-1} \ \& \ s_{n-1} \leq k) \right) \delta_{s_n,s_{n-1}} \right],
	\label{eq:priorsAbsorb}
\end{equation} 
where the indicator function $I(x_n \neq x_{n-1} \ \& \ s_{n-1} \leq k)$ is one only when both $x_n \neq x_{n-1}$  and $s_{n-1} \leq k$ are true. Notice that the above is an inhomogeneous chain having two modes: the first when the segment counting proceeds normally and the second when counting stops once the absorbing state is visited.  The $k$-segment problems for the event $s_N > k$ are then solved by using the above chain and clamping $s_N$ to the value $k+1$.  

The augmentation with counting variables results in a new HMM having the pair $(s_n,x_n)$ as the new extended state variable.  Given that $s_N=k$, so that any pair $(s_n,x_n)$ can jointly take at most $k M$ values, we can use the  Viterbi algorithm to obtain the MAP of $p(\bfx|\bfy, s_N = k)$, the forward pass of the F-B algorithm to obtain $p(s_N=k, \bfy)$ and the FF-BS algorithm to draw an independent sample from $p(\bfx|\bfy, s_N = k)$. A naive implementation of these algorithms can be 
done in $O(k^2 M^2 N)$ time. However,  this complexity can be further reduced to $O(k M^2 N)$ by taking into account the deterministic structure of the counting chain as discussed in the following section.    

\subsection{Efficient computation via dynamic programming \label{sec:dynprog}}

{\bf Optimal decoding.} We first describe the $k$-segment equivalent of the Viterbi algorithm for the optimal decoding problem under $k$-segment constraints, i.e.\ for obtaining the MAP of $p(\bfx|\bfy, s_N = k)$. This algorithm will be able to solve at once all such problems from $k=1$ up to a maximum $k=k_{max}$ by applying a single forward pass for the maximum value $k_{max}$ which requires $O(k_{max} M^2 N)$ operations. Then, by applying $k_{max}$ backtracking operations, each scaling as $O(N)$, we can obtain all $k_{max}$ optimal segmentations overall in $O(k_{max} M^2 N)$ time. 

More precisely, the Viterbi algorithm applies a forward pass where recursively $p(\bfy|\bfx)p(\bfx) p(\bfs_{\setminus N}, s_N=k_{max}|\bfx)$ is maximized with respect to the pair $(s_{n-1},x_{n-1})$ for any value of the next pair $(s_n,x_n)$. This can be implemented as a propagation of a message, which is a $k_{max} M $ dimensional vector,  as follows. The message is initialized to   
\begin{equation}
	\gamma(x_1,s_1) = \log p(y_1|x_1) +  \log p(x_1)  + \log p(s_1|x_1),
\end{equation}
which equals  $\log p(y_1|x_1) +  \log p(x_1)$ when $s_1=1$ and $-\infty$ when $s_1 > 1$. This message  
then is propagated recursively according to
\begin{equation}
	\gamma(x_n, s_n) = \log p(y_n|x_n) + \max_{x_{n-1},s_{n-1}} \left[ \gamma(x_{n-1}, s_{n-1}) + \log p(x_n|x_{n-1}) p(s_n|s_{n-1},x_{n},x_{n-1}) \right]. 
	\label{eq:gammamessage}
\end{equation}
\begin{equation}
	\delta(x_n, s_n) = (x_{n-1}^*,s_{n-1}^*),
	\label{eq:deltamessage}
\end{equation}
where the auxiliary message $\delta(x_n, s_n)$ simply stores the pair $(x_{n-1},s_{n-1})$ that gives the maximum in  (\ref{eq:gammamessage}) needed later in backtracking. Naively, the $n$th recursive update can be implemented in $O(k_{max}^2 M^2)$ time since each $s_n$ takes at most $k_{max}$ values and each $x_n$ takes $M$ values. However, for any given configuration of $(s_n,x_n)$ (out of the $k_{max} M$ possible), the permissible values for $s_{n-1}$ are either $s_{n-1} = s_n$ when $x_{n-1} = x_n$ or $s_{n-1} = s_n - 1$ when $x_{n-1} \neq x_n$. For all remaining configurations, $\log p(s_n|s_{n-1},x_{n},x_{n-1})= -\infty$,  so that these configurations need not to be checked when maximizing over $(s_{n-1},x_{n-1})$ for a certain pair $(s_n,x_n)$. Thus, the maximization in (\ref{eq:gammamessage}) can be done in $M$ operations resulting in $k_{max} M^2$ operations for the whole $n$th update. Subsequently, the full forward pass requires $O(k_{max} M^2 N)$ operations. Once the forward pass is completed, we have the final message $\gamma(x_N,s_N)$  (together with all auxiliary $\delta$ messages) from which we can obtain all $k_{max}$ optimal segmentations using backtracking as follows. For $k=1,\ldots,k_{max}$, we first compute 
\begin{equation}
	x_N^* = \arg \max_{x_N} \gamma(x_N,s_N=k).
	\label{eq:backtrack}
\end{equation} 
Then, starting from $(x_N^*,s_N^*=k)$ we backtrack recursively according to $(x_{n-1}^*,s_{n-1}^*) \leftarrow \delta(x_n^*,s_n^*)$ that recovers the optimal hidden path $\bfx^*$ having exactly $k$ segments. Each backtracking requires $O(N)$ simple indexing operations. 

{\bf Probability computation.} For the probability computation problem, we work similarly to the above Viterbi algorithm and we compute all joint densities $p(s_N=k,\bfy)$ for $k=1$ up to $k_{max}$ using the forward pass of the F-B algorithm applied to the augmented HMM. This recursively sums out each pair $(s_{n-1},x_{n-1})$ for any value of the next pair $(s_n,x_n)$, essentially passing through the so-called $\alpha$ message \citep{Bishop:2006}. This message is a $k_{max} M$ dimensional vector taking as initial value 
\begin{equation}
	\alpha(x_1,s_1) = p(y_1|x_1) p(x_1) p(s_1|x_1), 
\end{equation}
which equals  $p(y_1|x_1)p(x_1)$ when $s_1=1$ and $0$ otherwise. Then, the message is propagated according to the standard $\alpha$ recursion
\begin{equation}
	\alpha(x_n, s_n) = p(y_n|x_n) \sum_{x_{n-1}, s_{n-1}} \alpha(x_{n-1}, s_{n-1}) p(x_n|x_{n-1}) p(s_n|s_{n-1}, x_{n}, x_{n-1}). 
\end{equation}
This recursion scales as $O(k_{max} M^2)$ since the summation over $(x_{n-1}, s_{n-1})$ can be done in $O(M)$ time by taking advantage the  structure of the counting conditional $p(s_n|s_{n-1},x_{n},x_{n-1})$. As in any $\alpha$ recursion in a HMM, $\alpha(x_n,s_n)$ equals the density $p(x_n,s_n, y_1,\ldots,y_n)$ so that the final message is $\alpha(x_N,s_N) = p(x_N,s_N, \bfy)$, from which we can easily obtain 
\begin{equation}
	p(s_N=k,\bfy) = \sum_{x_N} \alpha(x_N,s_N=k),
\end{equation}   
for $k=1,\ldots,k_{max}$. Clearly, since  the computation of a single recursion of the $\alpha$ message takes $O(k_{max} M^2)$ time, the above computations require overall $O( k_{max} M^2 N)$ time. Given that the joint density $p(s_N=k,\bfy)$ has been obtained, we can compute exactly the posterior probability $p(s_N=k|\bfy)$ by dividing with the normalization constant $p(\bfy)$ (i.e.\ the overall likelihood of the HMM) obtained from the standard forward pass. 

Similarly to the above we can also define the so called backwards or $\beta$ message in the extended state-space HMM. 
Such message is useful when applying the EM algorithm for learning an HMM under $k$-segments constraints and its computation
will be described in section \ref{eq:em}.

{\bf Path sampling.} We now turn into the sampling problem where we wish to draw a path from the conditional $p(\bfx|s_N=k,\bfy)$. Such a path can be obtained by sampling a pair $(\bfx,\bfs_{\setminus N})$ from $p(\bfx,\bfs_{\setminus N}|s_N=k,\bfy)$ and then discarding $\bfs_{\setminus N}$. We apply the FF-BS algorithm that is based on the following decomposition 
\begin{equation}
	p(\bfx,\bfs_{\setminus N} | s_N=k, \bfy) = p(x_N | s_N=k, \bfy) \prod_{n=N-1}^1 p( x_{n}, s_n | x_{n+1}, s_{n+1}, y_1,\ldots,y_n),
\end{equation}
where the index $n$ in $\prod_{n=N-1}^1$ starts from $N-1$ and decrements down to one. Applying first the forward pass described above we have the final message $\alpha(x_N,s_N,\bfy)$ from which we can sample $x_N$ from $p(x_N | s_N=k, \bfy) \propto \alpha(x_N,s_N=k,\bfy)$.  Then, recursively we go backwards and each time we sample $(x_n, s_n)$, given the already sampled value of $(x_{n+1}, s_{n+1})$, from 
\begin{equation}
	p( x_{n}, s_n | x_{n+1}, s_{n+1}, y_1,\dots,y_n)  \propto  p(x_{n+1} | x_n ) p(s_{n+1} | s_n, x_{n+1}, x_n )  \alpha(x_n,s_n),
\end{equation}
where the message $\alpha(x_n,s_n) = p(x_n,s_n, y_1,\ldots,y_n)$ is known from the forward pass. Each sampling step takes $O(M)$ time (again due to the deterministic nature of the conditional $p(s_{n+1} | s_n, x_{n+1}, x_n))$ and the whole backward sampling requires $O(M N)$ time. If we wish to simultaneously sample from all conditional distributions $p(\bfx|s_N=k,\bfy)$, with $k=1,\ldots,k_{max}$, we can do this using a single forward pass that scales as $O(k_{max} M^2 N)$ and $k_{max}$ backward sampling iterations scaling as $O(k_{max} M N)$, so the overall complexity is $O(k_{max} M^2 N)$. 

Furthermore, very simple and straightforward modifications of the above procedures can deal with the more general constraint $k_1 \leq s_N \leq k_2$, where $1 \leq k_1 < k_2 \leq N$. For instance, if we wish to sample a path from  $p(\bfx| k_1 \leq s_N \leq k_2,\bfy)$, we need to first apply the forward pass for $k_{max}=k_2$ and then perform backwards sampling exactly as described above with the only difference that initially we sample $(x_N, s_N)$ from $p(x_N, s_N| k_1 \leq s_N \leq k_2, \bfy) \propto \alpha(x_N,s_N,\bfy) I(k_1 \leq s_N \leq k_2)$. Similarly, the $k$-segment inference problems associated with the special event $s_N > k$ can be efficiently  solved in $O((k+1)M^2 N)$ time by using the absorbing counting chain\footnote{An alternative is to assume the standard counting chain along with the event $k < s_N \leq N$. However, such a solution is very inefficient as it scales as $O(M^2 N^2)$ since $k_{max}$ must be chosen to be equal to $N$.} from (\ref{eq:priorsAbsorb}) and then applying exactly the above algorithms by clamping $s_N=k+1$. 

Finally, it is important to notice that running $k$-segment inference  up to some $k_{max}$ and setting $k_{max}+1$ as the absorbing state always gives a global summary of the posterior distribution that is guaranteed to be at least as informative as the standard Viterbi MAP path. More precisely, the events $c_{\bfx}=1,\ldots,c_{\bfx}=k_{max}$ and $c_{\bfx}>k_{max}$ comprise exclusive events that make up the whole set of paths for any value of $k_{max}$. Therefore, the probabilities  $p(c_{\bfx}=1|\bfy),\ldots,p(c_{\bfx}=k_{max}|\bfy)$ and $p(c_{\bfx}>k_{max}|\bfy)$, computed based on the forward pass in the augmented HMM, always sum up to one, while the set of the corresponding $k_{max}+1$ optimal paths must include the standard Viterbi MAP path, which will be either one of the paths from $1$ up to $k_{max}$ or the path with more segments than $k_{max}$. We refer to the above combined  sets of probabilities and optimal paths as the $k_{max}+1$ summary of the posterior distribution.

\subsection{Illustrative Example \label{sec:graphillustr}}

Here, we give a graphical illustration of optimal decoding and path sampling under $k$-segment constraints. For this, 
we simulated a data sequence according to $y_n | x_n, \bfm, \sigma^2 \sim \mathcal{N}(m_{x_n}, \sigma^2 ),~n=1,\dots,N=1000$, where the 
hidden sequence $\bfx = \{x_n\}_{n=1}^N$ was given by a Markov chain with $M=3$ states, $\bfm = \{ -2, -1, 1 \}$ and $\sigma= 0.9$. 

\begin{figure}[!h]
\centering
\begin{tabular}{c}
{\includegraphics[scale=0.2]{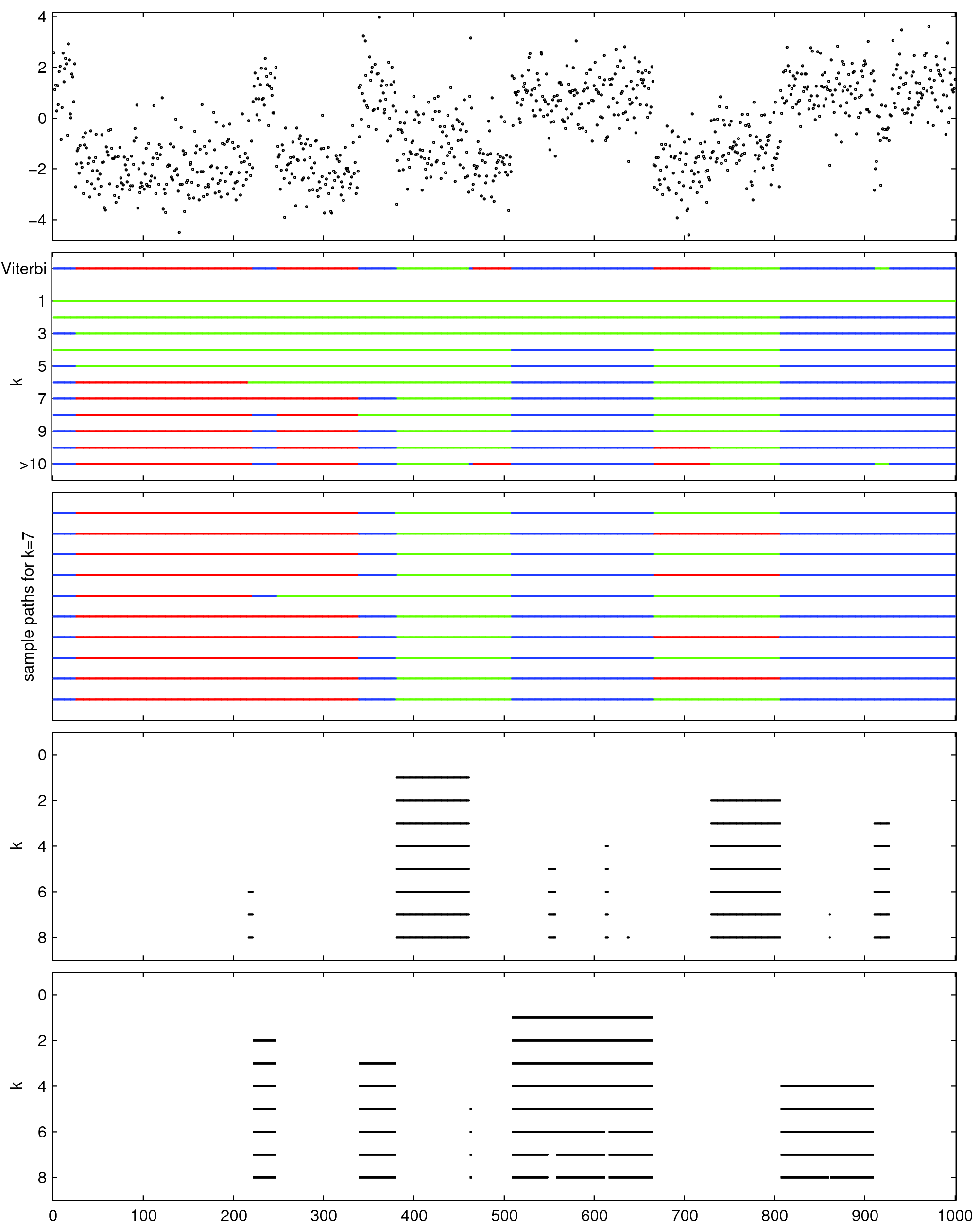}}
\end{tabular}
\caption{The panel in the first row shows the simulated data sequence. The panel in the second plot displays the 
standard Viterbi path and $11$ optimal paths corresponding to the $k_{max}+1$ summary (with $k_{max}=10$) of the $k$-segment inference. 
Each path is depicted with the three states shown in different colors (the state with emission Gaussian density of ground-truth mean $-2$
is shown in red, the one with mean $-1$ shown in green and the third one with mean $1$ shown in  blue). The panel in the third row shows $10$ sample paths
 obtained by the FF-BS algorithm under the constraint that exactly $7$ segments occur. The panel in the forth row 
illustrates generalized counting (section \ref{sec:countingsubset}) so that the optimal paths having $0$ up to $8$ segments from the second  
state are shown. For clarity, only the segments of the second state of interest are displayed with black solid lines. 
The final panel in the last row illustrates counting excursions (section \ref{sec:excursions}) having as the null set the first and second 
states while the third one is the single state in the abnormal set.  Again for clarity, only the excursion segments (more precisely only 
the abnormal sub-segment, i.e.\ excluding the start and end points) are shown using black solid lines.}
\label{fig:illustrativePlots}
\end{figure}

Using the simulated data, shown in the first row of Figure \ref{fig:illustrativePlots}, we fitted a three-state HMM using the EM algorithm which recovered parameter estimates very close to the ground-truth ones. We then computed the standard Viterbi path and obtained the optimal segmentations, associated with the $k_{max}+1$ summary, using the $k$-segment equivalent of the Viterbi algorithm with $k_{max}=10$. These are shown in the second row of Figure \ref{fig:illustrativePlots}. 

Each such path is displayed so that the three states are shown with different color. On top, the Viterbi path is displayed, containing $14$ segments, and then the $11$ paths of the $k_{max}+1$ summary. The first $10$ paths of the latter summary provide a coarse-to-fine hierarchical segmentation of the data sequence where the number of segments increase by one each time. Notice that two consecutive segmentations, do not always follow the principle used in circular binary segmentation algorithm \citep{cbinarysegm2004}, i.e.\ the $k+1$th segmentation might not be obtained by splitting into two segments a single segment from the $k$th one. Such a latter approach is sub-optimal. Also, notice that the final path that corresponds to the absorbing state (labelled with $>10$ in the figure) is precisely the standard Viterbi path. The third panel of Figure \ref{fig:illustrativePlots} illustrates path sampling under $k$-segment constraints using the FF-BS algorithm in the augmented HMM. In particular, $10$ samples are shown that are constrained to have exactly $k=7$ segments. 

We remark that the application of our $k$-segment algorithms, so far, has been applied entirely retrospectively to an HMM fitted using a very standard and common approach in a simple but generic model set-up. The $k$-segment constraints are not involved in the model fitting process but are applied retrospectively to provide a rich exploration of the posterior sequence space where qualitatively diverse segmentations are reported. For the expenditure of some computational time, the application of $k$-segment generalizations for optimal decoding, probability computation and path sampling provides the HMM user with alternative summaries.

\section{$k$-segment inference in practice  \label{sec:statinference}}

So far we have presented novel recursions for HMM inference that are applied assuming a fixed value for the parameters $\bftheta$. In this section, we discuss how we could use these recursions in a general statistical estimation problem with HMMs. More precisely, we will analyze the following two uses of $k$-segment constraints: i) the retrospective or {\it a posteriori} use where the parameters of the HMMs have been fitted beforehand and $k$-segment inference is used as a meta-analysis tool (section \ref{sec:aposteriori}) and ii) the prospective or {\it a priori}  use where the constraints are introduced during model fitting so that they actively influence the model parameters (section \ref{sec:kseglearning}). 

\subsection{Retrospective utility of $k$-segment constraints \label{sec:aposteriori}}

The most obvious practical use of $k$-segment inference is the following. Given an HMM with fixed parameters, e.g.\ estimated by ML training, apply the recursions of Section \ref{sec:dynprog} to explore the posterior distribution over the hidden sequences. Some questions that arise are: i) what is justification of this approach? ii) when is it suitable? and iii)  how can be extended in a Bayesian estimation setting? 
 
To start with the first question, a way to formalize the {\it a posteriori} use  of $k$-segment inference is using decision-theoretic arguments where actions are taken {\it a posteriori} given that beliefs about a system are described by some fixed and known probability model. For instance, the computation of the MAP hidden path $\bfx^*$ of $p(\bfx|c_{\bfx}=k,\bftheta_{ML},\bfy)$, where $\bftheta_{ML}$ is some value obtained by ML training, can be considered as choosing the action $\bfz$ that maximizes the following expected utility 
\begin{equation}
	\bfx^* = \arg \max_{\bfz} \sum_{\bfx}  u(\bfz,\bfx; c_{\bfx} = k) p(\bfx|\bftheta_{ML}, \bfy), 
\end{equation} 
where the $0$-$1$ utility function $u(\bfz,\bfx; c_{\bfx} = k)$ takes the value one  only when both $\bfz=\bfx$ and $\bfx \in \{\bfx|c_{\bfx}=k\}$. The introduction of constraints using the counting auxiliary variables allows for efficient maximization of the above expected utility using dynamic programming.

Regarding the second question, notice that the assumption of the {\it a posteriori} use of $k$-segment constraints implies that the model parameters, which quantify the structure of the hidden states, are inferred independently of whatever $k$-segment constraints we may wish to consider. In practice, this is sensible when the hidden states are clearly interpretable classes so that the emission densities model true class conditional densities and the transition matrix represents meaningful spatial correlation between these classes.

In a Bayesian setting, a prior distribution $p(\bftheta)$ is placed on the HMM parameters and then it is updated to a posterior distribution $p(\bftheta|\bfy)$ by conditioning on the observed data. Following similar arguments with the ones above,  under a posteriori use of $k$-segment constraints we assume that model parameters $\bftheta$ are conditionally independent from any constraint, say $c_{\bfx}=k$, given the observed data $\bfy$, i.e.\  $p(\bftheta|c_{\bfx}=k,\bfy) = p(\bftheta|\bfy)$.

Again such an assumption is sensible when the parameters describe the structure of true classes which is independent of any constraint in the classification procedure. Assuming now that computationally the posterior distribution $p(\bftheta|\bfy)$ is realized by a set of samples $\{\bftheta^{(t)}\}_{t=1}^T$, obtained say by some MCMC algorithm, the three $k$-segment inference problems can be tackled as follows. Firstly, the computation of $p(c_{\bfx}=k|\bfy)$ can be done according to 
\begin{eqnarray}
	p(c_{\bfx} = k |\bfy) 	& = & \int \sum_{\bfx}  I(c_{\bfx} = k ) p(\bfx|\bftheta, \bfy) p(\bftheta|\bfy) d \bftheta \nonumber \\ 
							& = & \int p(c_{\bfx} = k | \bftheta, \bfy) p(\bftheta | \bfy)  d \bftheta \nonumber \\ 
							& \approx & \frac{1}{T} \sum_{t=1}^T p(c_{\bfx} = k | \bftheta^{(t)}, \bfy),
\end{eqnarray}
where the Rao-Blackwellization when summing out $\bfx$ is carried out by the forward recursion of the F-B algorithm in the augmented HMM with the final counting variable clamped to value $k$. More precisely, for each parameter sample $\bftheta^{(t)}$, this recursion computes the probability $p(c_{\bfx} = k, \bfy| \bftheta^{(t)})$ in $O( k M^2 N)$ time from which we obtain $p(c_{\bfx} = k|\bftheta^{(t)}, \bfy)$ by normalizing with $p(\bfy|\bftheta^{(t)})$ obtained in $O(M^2 N)$ time using the standard forward pass in the unconstrained HMM. Similarly, to find the MAP of $p(\bfx| c_x=k, \bfy)$ we are based on the expression
\begin{eqnarray}
	p(\bfx| c_x=k, \bfy) 	& = & \int p(\bfx|\bftheta, c_x=k,\bfy)  p(\bftheta|c_x=k,\bfy)  d \bftheta \nonumber \\
							& = & \int p(\bfx|\bftheta, c_x=k,\bfy)  p(\bftheta|\bfy)  d \bftheta, 
	\label{eq:px_bayesian}
\end{eqnarray}
where we used the conditional independence assumption. Similarly to the unconstrained MAP of $p(\bfx|\bfy)$ \citep{Scott2002}, the above cannot be maximized analytically with respect to $\bfx$ (notice also that applying Monte Carlo in the final integral will not help) and therefore we consider an approximate solution of the form
\begin{equation}
	\hat{\bfx} = \arg \max_{\bfx} p(\bfx|c_x=k,\hat{\bftheta},\bfy),   
\end{equation}
obtained by applying the Viterbi algorithm in the augmented HMM. Here, $\hat{\bftheta}$ is an approximation of the MAP 
of $p(\bftheta|\bfy)$ computed from  
the samples, i.e.\  $\hat{\bftheta} = \bftheta^{(t^*)}$, $t^* = \arg \max_{1 \leq t \leq T} \left[ p(\bfy |\bftheta^{(t)}) p(\bftheta^{(t)}) \right]$.
Regarding path sampling, if we wish to draw a path $\bfx$ from $p(\bfx| c_x=k,\bfy)$, then by following  eq.\ (\ref{eq:px_bayesian}) we can choose a parameter sample $\bftheta^{(t)}$ uniformly form the set $\{\bftheta^{(t)}\}_{t=1}^T$, and then draw a path from $p(\bfx|\bftheta^{(t)}, c_x=k,\bfy)$ using the FF-BS algorithm in the augmented HMM. 

Finally, it is worth discussing how an HMM with its hidden states being true classes can be fitted to data in practice. It is important to note that this cannot be done in a completely unsupervised manner, as in such case the classes are not identifiable. Clearly, we need to inject knowledge into the model about which state corresponds to which class. One extreme way to achieve this is to use completely supervised learning so that the data consist of both the sequence $\bfy$ and the class-label sequence $\bfx$, i.e.\ $\bfx$  is fully observed. ML training in such case simplifies significantly and the standard EM algorithm for the HMM is not needed. A similar scenario, somehow more realistic, is to use semi-supervised learning where the path $\bfx$ is partially observed. Such supervised or semi-supervised  approaches can be also implemented in two phases where in the first  phase some of the class conditional densities are found using labelled data and then they are used as fixed transition densities for segmental classification within a HMM. We will make use of such latter approach to train a HMM for text retrieval in Section \ref{sec:topics}. A different way is to inject knowledge about the classes is via the prior $p(\bftheta)$ by following a fully Bayesian approach or penalized ML. For instance, the classes might have a natural ordering or pairwise proximity which could be taken into account by choosing a suitable prior $p(\bftheta)$. This will resolve the identifiability issues by assigning the hidden states to classes implicitly via the prior while otherwise training could be performed in unsupervised manner with the path $\bfx$ being fully unobserved.

\subsection{Learning with $k$-segment constraints  \label{sec:kseglearning}}

A second way to use $k$-segment constraints is to incorporate them in the model  fitting process so that the inferred parameters, and hence the structure of the hidden states, will depend on these constraints. In contrast to the hidden states being classes, here the hidden states are {\em latent variables} that simply add flexibility in fitting the data, similarly for instance to latent components in unsupervised mixture density estimation, A suitable application is optimal compression of a data sequence by minimizing the reconstruction error, or equivalently maximizing model fitting, between the observed sequence and the representation provided by the model. In such case a $k$-segment constraint could represent a fixed budget in the reconstruction process and clearly it would be more efficient to incorporate the constraint during the model fitting.  

To this end, next we describe the technical details of using $k$-segment constraints for learning an HMM either by applying the EM algorithm (section \ref{eq:em}) or by applying Bayesian approaches (section \ref{sec:bayes}).

\subsubsection{Expectation-Maximization \label{eq:em}}

Suppose that together with the data $\bfy$ we have an additional piece of information about the number of segments in the hidden path. Specifically, we shall assume that this number cannot be larger than $k$ while any other constraint,  such as being exactly $k$, can be dealt with in a similar manner. By incorporating this constraint in the augmented HMM we obtain the following joint density:   
\begin{equation}
	p(\bfy,\bfx,\bfs) = p(\bfy|\bfx) p(\bfx) p(s_N \leq k, \bfs_{\setminus N}|\bfx),
	\label{eq:augmjoint}
\end{equation}
where the evidence $s_N \leq k$ reflects the information about the maximum number of segments allowed. Notice that incorporating the constraint simply amounts for constraining each counting variable $s_n$ to take the values $1,\ldots,k$.   

We would like now to apply the EM algorithm to learn the parameters $\bftheta$ for which we need to write down the auxiliary $Q$ function and subsequently derive the E and M steps. Since the factor $p(s_N \leq k, \bfs_{\setminus N}|\bfx)$ does not contain learnable parameters, the auxiliary $Q$ function can be written as 
\begin{equation}
	Q(\bftheta;\bftheta^{\text{old}}) = \mathbbm{E}_{p(\bfx|s_N \leq k,\bfy,\bftheta^{\text{old}})}[\log p(\bfy|\bfx,\bftheta) p(\bfx,\bftheta)] + \text{const},
\end{equation}
where $\bftheta^{\text{old}}$ denotes the current parameter values. This function has exactly the same form with the auxiliary function in the unconstrained HMM with the only difference being that $p(\bfx|\bfy,\bftheta^{\text{old}})$ is replaced by $p(\bfx|s_N \leq k,\bfy,\bftheta^{\text{old}})$.
 
The E step simplifies to computing all marginals  $p(x_n|s_N \leq k,\bfy,\bftheta^{\text{old}})$ and all pair-wise marginals $p(x_{n-1},x_n|s_N \leq k,\bfy,\bftheta^{\text{old}})$ which can be obtained by applying  the F-B algorithm in the augmented HMM. Given the current $\bftheta^{\text{old}}$ (omitted next for brevity), this algorithm  computes the $\alpha$ messages, as shown in section \ref{sec:dynprog}, and the backward or $\beta$ messages, so that the first $\beta$ message is initialized to 
unity (i.e.\ $\beta(x_N,s_N)=1$) and subsequent $\beta$ messages are recursively obtained according to   
\begin{equation}
	\beta(x_n, s_n) =  \sum_{x_{n+1}, s_{n+1}} \beta(x_{n+1}, s_{n+1}) p(y_{n+1}|x_{n+1}) p(x_{n+1}|x_n) p(s_{n+1}|s_n, x_{n+1}, x_n). 
\end{equation}
Given that each $s_n$ takes $k$ values, the $\beta$ messages are computed in overall  $O(k M^2 N)$ time. Having stored all $\alpha$ and $\beta$ messages the desired marginals and pair-wise marginals   are obtained from
\begin{equation}
	p(x_n|s_N \leq k, \bfy) \propto \sum_{s_n=1}^k \alpha(x_n,s_n) \beta(x_n,s_n),
	\label{eq:ksegmarginal} 
\end{equation}
\begin{equation}
	p(x_{n-1}, x_n|s_N \leq k, \bfy) \propto \sum_{s_{n-1}, s_n=1}^k \alpha(x_{n-1},s_{n-1}) p(y_n|x_n) p(x_n|x_{n-1}) p(s_n|s_{n-1},x_n,x_{n-1}) \beta(x_n,s_n),
	\label{eq:ksegpairmarg} 
\end{equation}
which involve summing out the auxiliary counting variables. Given these quantities from the E step, the form of M step remains the same as in
 unconstrained HMMs. The iteration between the above E and M steps leads to a local minimum of the likelihood $p(c_{\bfx}\leq k,\bfy)$. 

Further, as mentioned earlier, deriving EM algorithms under similar constraints can be done as above. For instance, if we wish to apply the EM algorithm by assuming the number of segments to be exactly equal to $k$, we need to clamp the final counting variable $s_N$ to the value $k$. 

\begin{figure}
\begin{center}
\begin{tabular}{cc}
{\includegraphics[scale=0.11]
{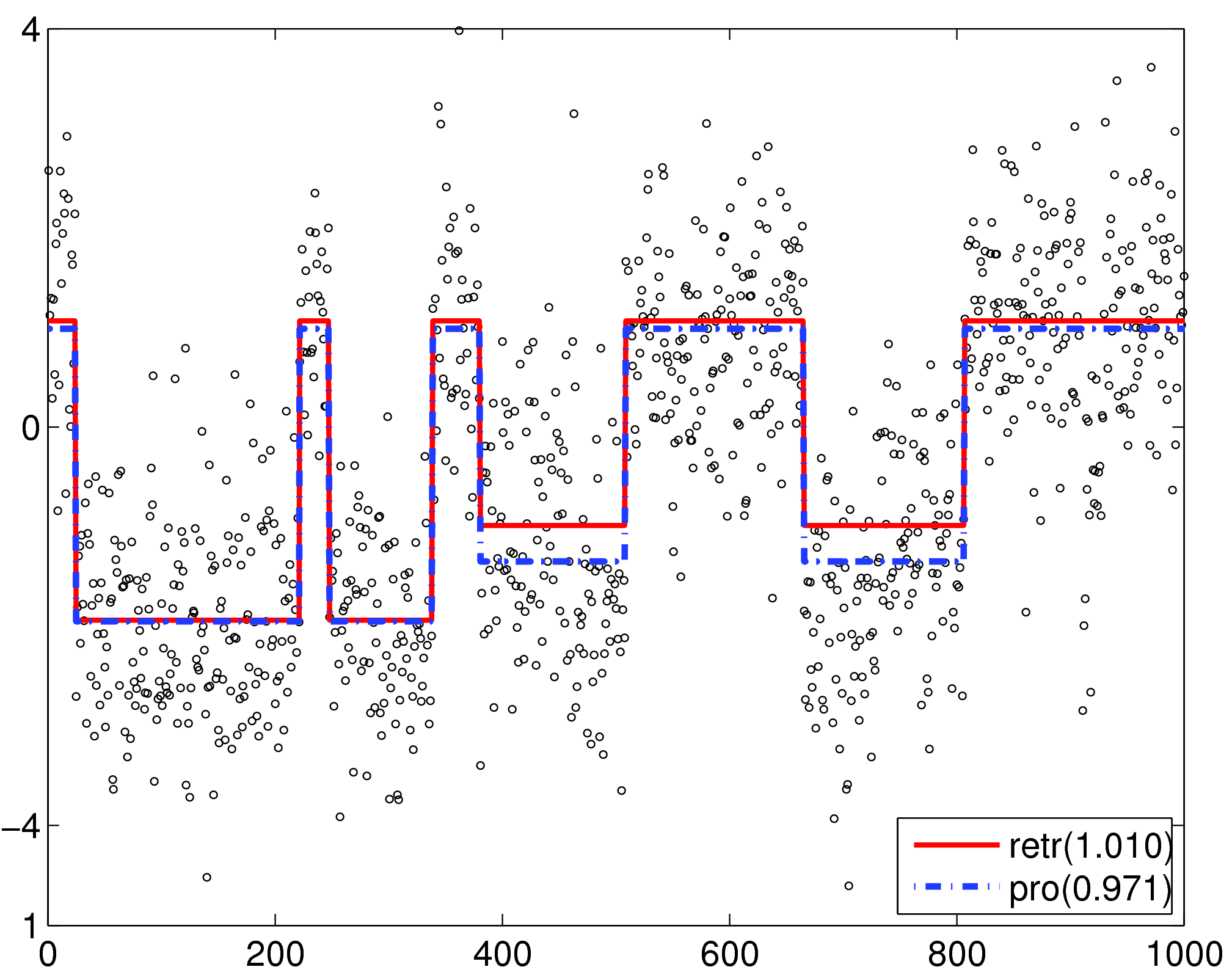}} & 
{\includegraphics[scale=0.11]
{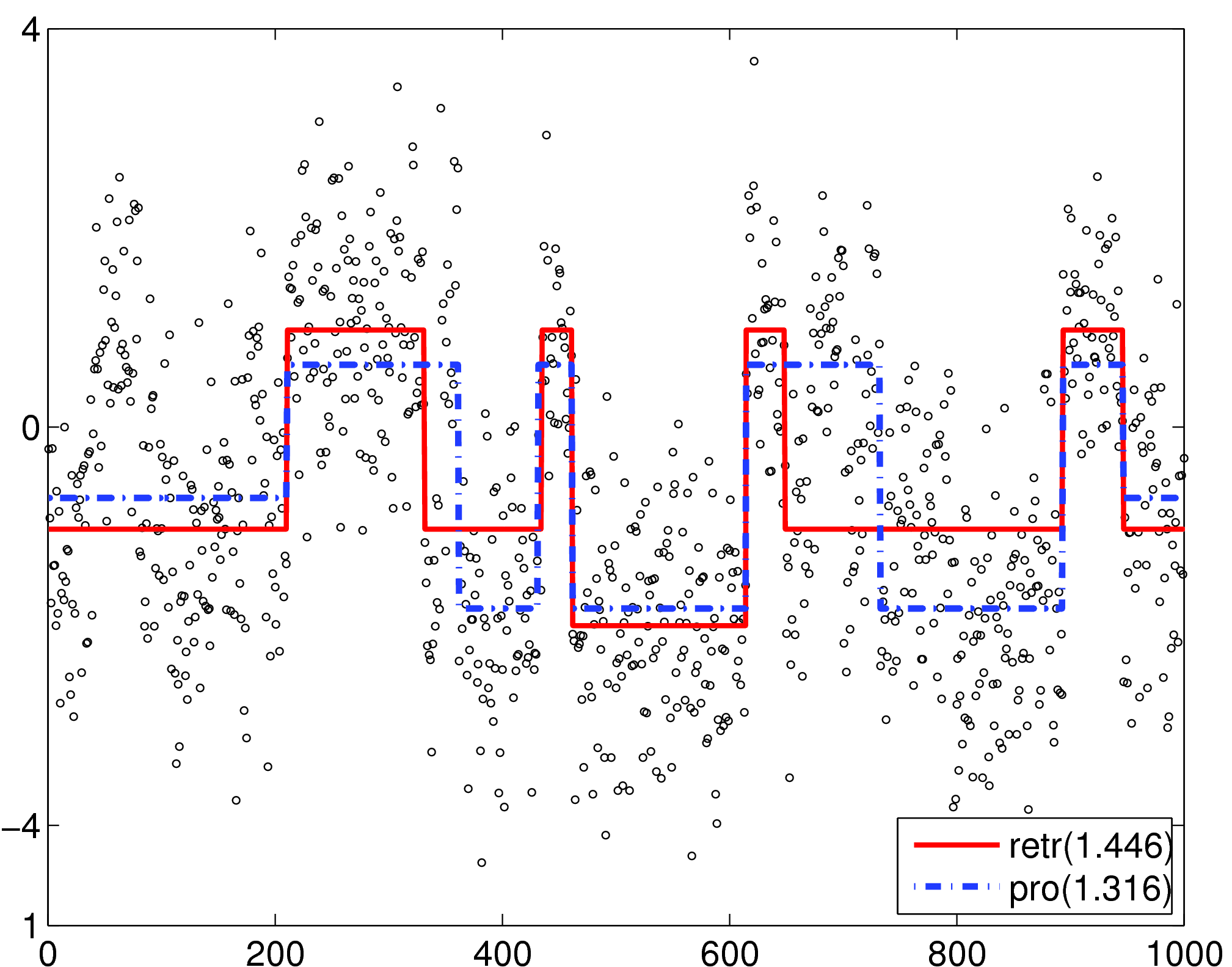}} \\ 
{\includegraphics[scale=0.11]
{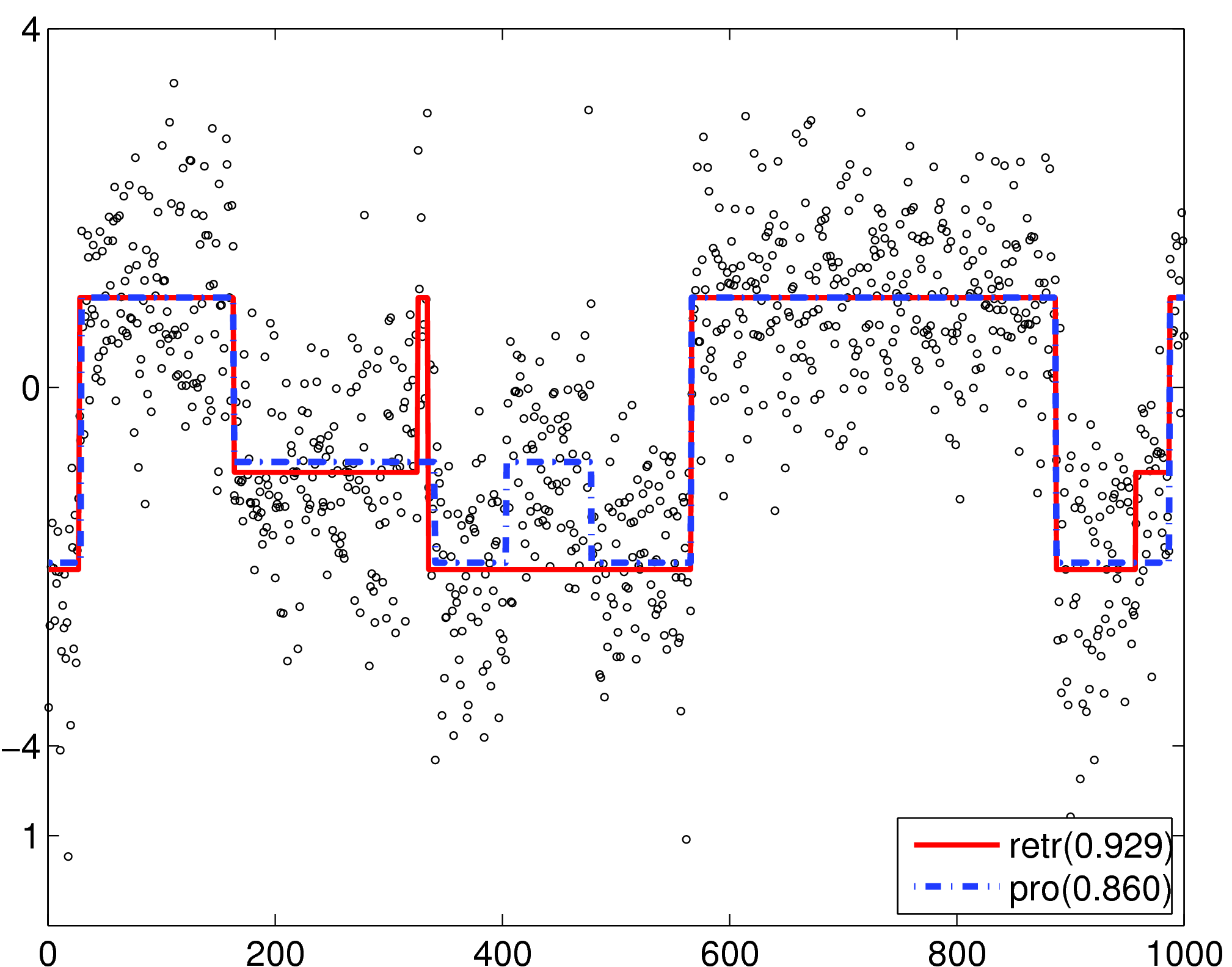}} &
{\includegraphics[scale=0.11]
{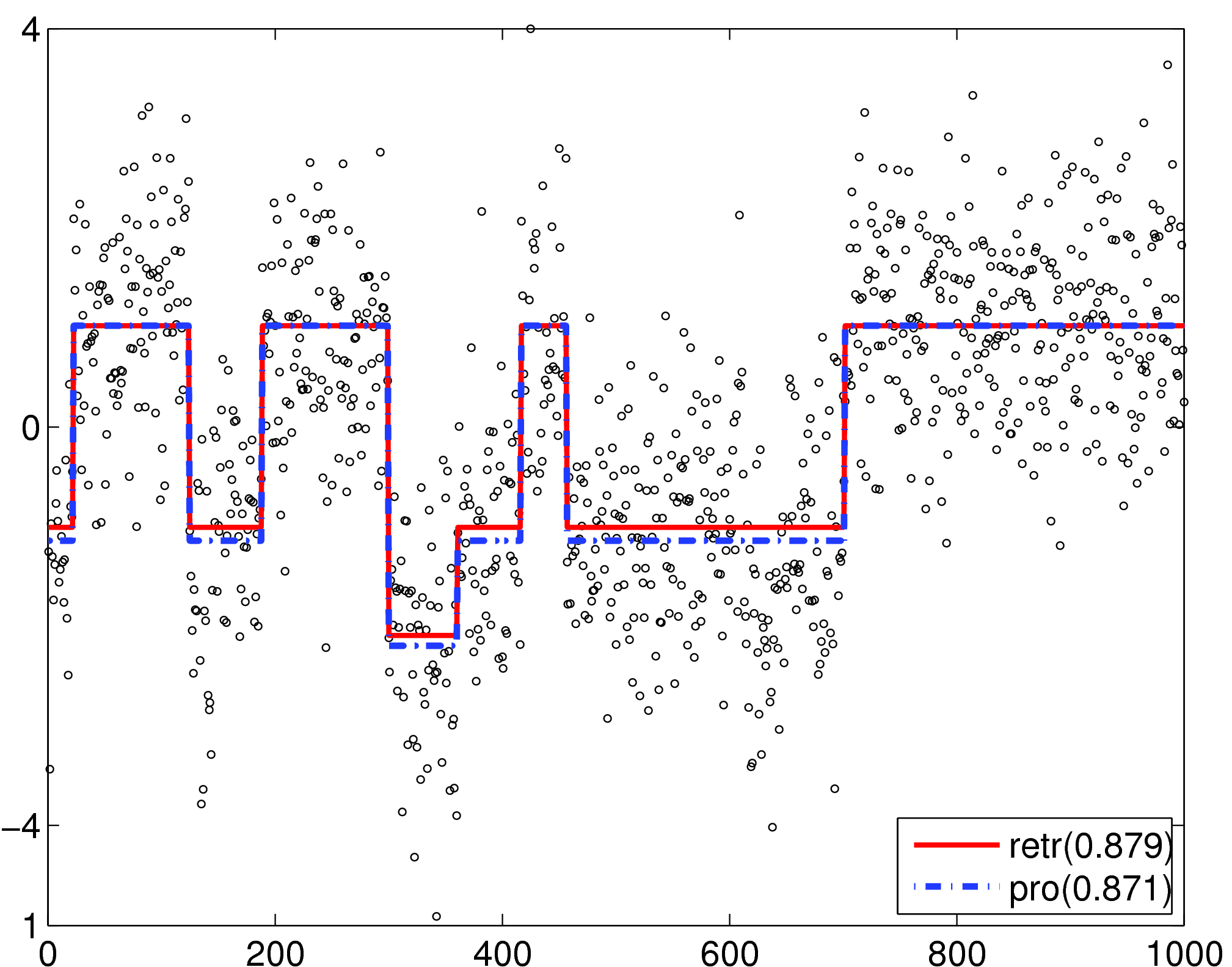}} \\ 
{\includegraphics[scale=0.11]
{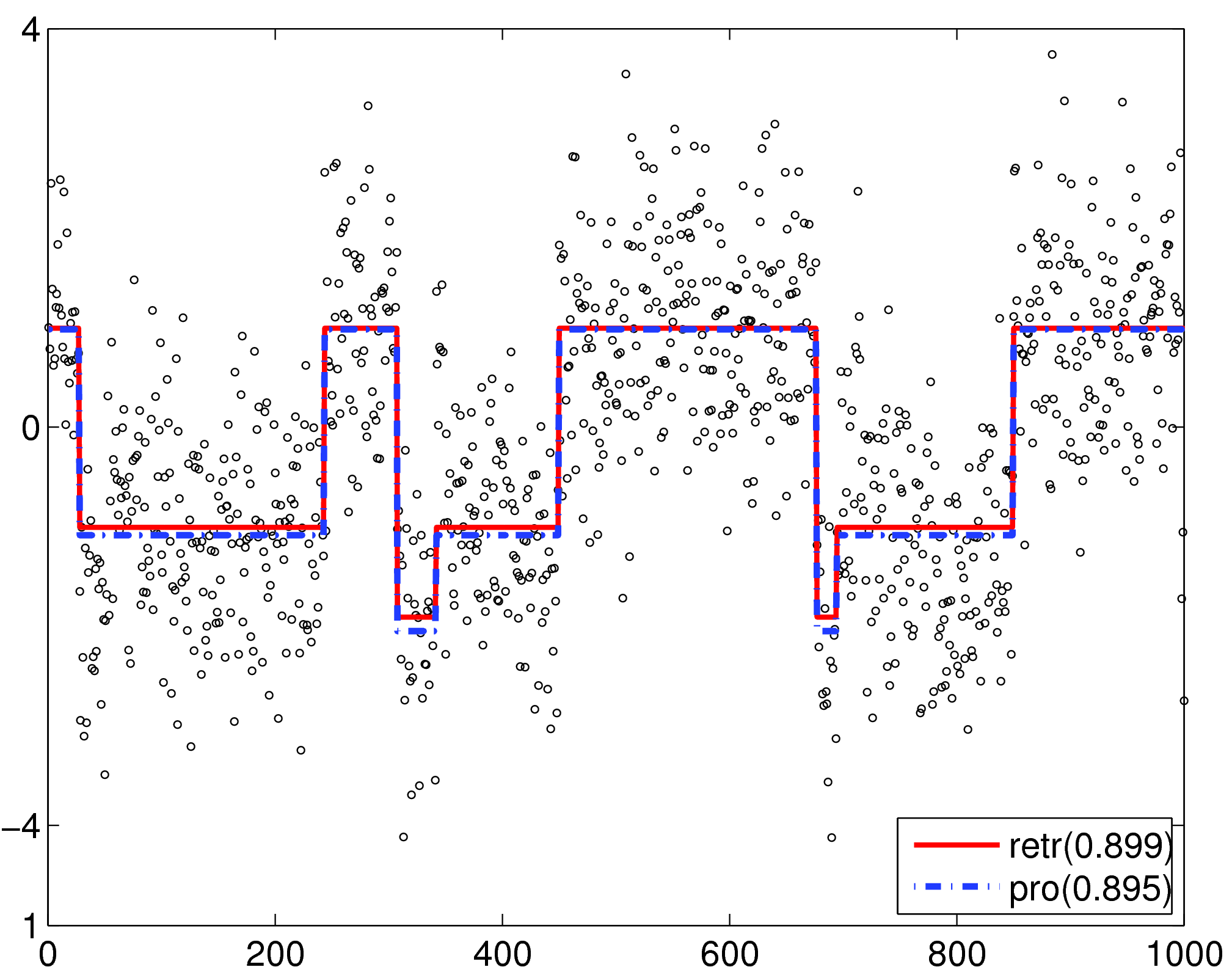}} & 
{\includegraphics[scale=0.11]
{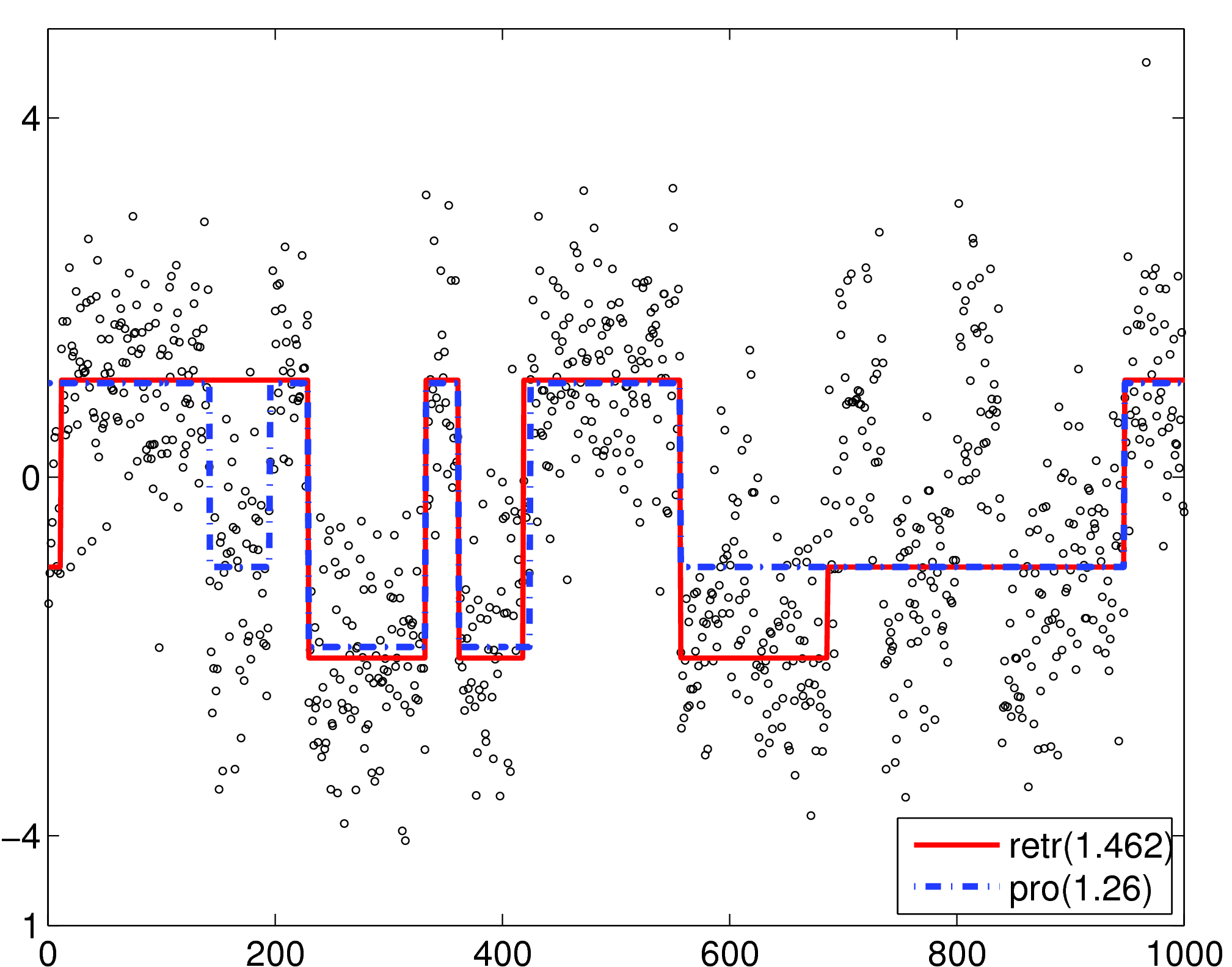}} 
\end{tabular}
\caption{Each of the six panels corresponds to a different simulated sequence. In each panel, 
the prospective (blue dashed line) and the retrospective (red solid line) optimal paths, under the $k$-segment constraint $c_{\bfx} \leq 9$, 
are displayed together with the data. The paths are shown as piece-wise constant functions 
formed by the means of the Gaussian emission densities. MSEs of the two reconstructions are given in the 
legends of the panels inside parentheses.}
\label{fig:em}
\end{center}
\end{figure}

To give an example of using the above learning algorithms, we consider six simulated sequences (included the one from Figure  
\ref{fig:illustrativePlots}) and we apply EM under the $k$-segment constraint $c_{\bfx} \leq 9$.
The six panels in Figure \ref{fig:em} shows the optimal $k$-segment paths (blue dashed lines) 
obtained after having optimized the HMM with the constrained EM described above. The corresponding 
retrospective optimal paths associated with the same constraint (red solid lines), i.e.\ the path found
 after having optimized the HMM parameters using the 
standard unconstrained EM, are also displayed. Notice that each path is shown as a  
piece-wise constant function formed by the mean values of the Gaussian emission densities indicated by the states along the 
path. Each piece-wise constant function gives a reconstruction of the observed sequence. 
The mean squared errors (MSEs) for the retrospective and prospective reconstructions are shown inside parentheses in the legend 
of each panel. As expected, there MSEs indicate that incorporating a $k$-segment constraint during model fitting gives 
better reconstruction. In fact, by initializing the parameters
in the constrained EM from the final values obtained by the standard EM  
should always lead to a likelihood value  which is higher or equal 
to the corresponding value in the retrospective model. 

\subsubsection{Bayesian approaches \label{sec:bayes}}

It is also possible to learn an HMM under $k$-segment constraints using Bayesian inference and here we briefly outline
 how this can be done using Gibbs sampling. Consider a Bayesian HMM with a prior distribution $p(\bftheta)$ on the parameters and a joint density 
\begin{equation}
	p(\bfy,\bfx, \bfs, \bftheta) = p(\bfy|\bfx,\bftheta) p(\bfx|\bftheta) p(\bftheta) p(s_N \leq k, \bfs_{\setminus N}|\bfx).
	\label{eq:augmjoint}
\end{equation}
where, as in the previous section, we assumed that the number of segments cannot exceed $k$. Notice that, while $\bftheta$ and $\bfs$ are conditionally independent given $\bfx$,  marginally they are dependent because of the constraint $s_{N} \leq k$. We aim to compute the posterior distribution $p(\bfx, \bfs,\bftheta|s_N \leq k, \bfy)$ and since this is too expensive we resort to Gibbs-type of sampling where we iteratively sample the paths $(\bfx,\bfs)$ from the conditional $p(\bfx,\bfs|\bftheta, s_N \leq k, \bfy)$ and the parameters $\bftheta$ from $p(\bftheta| \bfx, \bfy)$. The first step corresponds precisely to the path sampling under a $k$-segment constraint presented in section \ref{sec:dynprog} using FF-BS in the augmented HMM. The second step requires simulating from the posterior conditional over parameters and clearly this will always be identical with the corresponding step when sampling in the unconstrained HMM. Also, when this step involves exact simulation from $p(\bftheta| \bfx, \bfy)$ the full algorithm is precisely Gibbs sampling, otherwise it is Metropolis-within-Gibbs where $\bftheta$ is sampled from a proposal distribution and then it is accepted or rejected.

\section{Extended $k$-segment inference problems \label{sec:extensions}}

In this section, we discuss extensions to the basic $k$-segment inference problems considered in section \ref{sec:theory}. Specifically, in section \ref{sec:countingsubset} we show how to solve generalized $k$-segment inference problems where we are interested in transitions of a particular type. In section \ref{sec:excursions}, we extend the framework in a different direction by showing how to extract highly non-Markovian events along the HMM hidden path which consist of excursions from null states to abnormal states. 

\subsection{Counting segments satisfying certain constraints \label{sec:countingsubset} }

In several applications of HMMs, we may wish to solve more general $k$-segment inference problems associated with probability events involving certain types of segments and transitions. For example, we could have a natural sub-group of states $\mathcal{A} \subset \{1, \ldots, M\}$ and we would  like to classify the observed sequence in terms of the occurrence or not of $\mathcal{A}$ based on the computation of the associated posterior probability. This problem consists of an example of generalized $k$-segment inference and in this section we show how this and related problems can be solved using auxiliary counting variables. 

In a hidden path of an HMM (assuming an irreducible transition matrix) we can encounter $M (M-1)$ possible transitions. We can denote this set of all transitions by an $M \times M$ binary matrix $C$ having ones everywhere and zeros in the diagonal, i.e.\ $C(i,j) = I(i \neq j)$. Such a matrix characterizes the standard $k$-segment inference problems described earlier where all segments are of interest and are all counted. When we care about a subset of transitions, we can modify $C$ so that $C(i,j) = 1$, if both $i \neq j$ and the transition $i \rightarrow j$ belongs to this subset. One way to visualise this is to think of colouring certain transitions in the HMM. Then, we will be interested in  counting segments generated from only those coloured transitions. Furthermore, in order to be flexible about the inclusion of the initial segment (which is not the result of a transition) in the probability event, we can define an $M$-dimensional binary vector $\bfmu$ indicating the subset of values of the initial state $x_1$ that are of interest. Then analogously to equation (\ref{eq:sxtrans}), we can define 
\begin{equation}
	c_{\bfx} = \mu(x_1) + \sum_{n=2}^N C(x_{n-1},x_n),
	\label{eq:sxtransGen}
\end{equation}
which denotes the number of segments along the hidden path $\bfx$ which are compatible with the constraints $(\bfmu,C)$. Subsequently, we can define probability events of the form $c_{\bfx} = k$, $k_1 \leq c_{\bfx} \leq k_2$, the special events $c_{\bfx} > k$ and etc, and subsequently formulate all associated $k$-segment inference problems as described in section \ref{sec:kseg}. 

To solve all these new problems, we introduce again auxiliary counting variables $\bfs$  and define a suitable counting Markov chain $p(\bfs|\bfx)$  that generates deterministically the variables in $\bfs$ given the path $\bfx$. This chain has the same structure with eq.\ (\ref{eq:priors}) but with the following modified conditionals: 
\begin{equation}
	p(s_1|x_1) = \mu(x_1) \delta_{s_1,1} + (1 - \mu(x_1) ) \delta_{s_1,0},
	\label{eq:newchain1}
\end{equation}
\begin{equation}
	p(s_n|s_{n-1},x_{n-1},x_n) = C(x_{n-1}, x_n) \delta_{s_n,s_{n-1}+1} + (1 - C(x_{n-1}, x_n )) \delta_{s_n,s_{n-1}}.
	\label{eq:newchain2}
\end{equation}
Here, $s_1$ is set to one only for the subset of values of $x_1$ compatible with $\bfmu$,  otherwise it remains zero and the associated initial segments are not counted.  The case of counting always the first segment corresponds to the special case where $\mu(x_1=i) = 1$, for each $i$, in which case $p(s_1|x_1)$ simplifies to $\delta_{s_1,1}$. Similarly,  the conditional $p(s_n|s_{n-1},x_{n-1},x_n)$ is such that $s_n$ increases only when $C(x_{n-1}, x_n)=1$ so that new segments for which $x_{n-1} \neq x_n$  and  $C(x_{n-1}, x_n)=0$ are not counted. Clearly, counting any segment is obtained as a special case for which $C(x_{n-1}, x_n)= I(x_{n-1} \neq x_n)$. 

All dynamic programming recursions of section \ref{sec:dynprog} are applicable to the above generalized $k$-segment inference problems.  Given that we solve these problems for $k=1$ up to $k=k_{max}$, the time complexity of these algorithms can be either  $O(k_{max} M^2 N)$ when the vector $\bfmu$ is equal to one everywhere or $O( (k_{max} +1) M^2 N)$ when some of the elements of $\bfmu$ are zero.  In the latter case the term $k_{max} + 1$ appears simply because each counting variable $s_n$ can take $k_{max}+1$ values.  Finally, standard Viterbi, F-B and FF-BS algorithms for HMMs are obtained as special cases of  generalized $k$-segment recursions corresponding to setting $\bfmu$ and $C$ to zero.   To make this clear, notice that in such case none of the segments along the path $\bfx$ are counted so that $k$ can take only the value zero, i.e.\ $p(c_{\bfx}=0|\bfy)=1$ and $p(\bfx|c_{\bfx}=0,\bfy) = p(\bfx | \bfy)$. Therefore, in such case the   generalized $k$-segment recursions reduce to the standard HMM recursions having complexity $O(M^2 N)$ which shows that the $k$-segment algorithms provide a more general inference methodology for HMMs.

Finally, to illustrate optimal decoding in a generalized $k$-segment setting, we consider again the 
simulated data of Figure \ref{fig:illustrativePlots}. Suppose, we would like to count segments from the second
(green) state only. The constraints $(\bfmu,C)$ we need to use are  $\bfmu=[0 \ 1 \ 0]$ and $C = [0 \ 1 \ 0; \ 0 \ 0 \ 0; \ 0 \ 1 \ 0]$ 
(where $;$ separates the rows of $C$). The fourth row of Figure \ref{fig:illustrativePlots} shows several optimal paths 
 having $0$ up to $8$ segments associated with counting the second state in the HMM. 

\subsection{Extracting excursions  using two layers of auxiliary variables \label{sec:excursions}}

In several applications of HMMs where hidden states correspond to true states of nature, there is often a subset of states (in the simplest case just a single 
state) considered as normal or null states while the remaining ones represent abnormalities. In such applications the practitioner might be interested to identify {\em excursions} where the hidden path moves from any null state to abnormal states and returns back to a null state. Extracting such events using a $k$-segment formulation is challenging because an excursion has a high order Markov structure and therefore it cannot be identified by just comparing two consecutive states. To this end, next we describe a generalization of our augmentation framework with counting variables that efficiently solves the excursion problem. 

We first give a precise definition of an excursion. Suppose in HMM the states are divided into two groups: the null set $\mathcal{N} \subset \{1,\ldots,M\}$ and the abnormal set
$\overline{\mathcal{N}}= \{1,\ldots,M\} \setminus \mathcal{N}$. An excursion is any  sub-path $(x_i,\ldots,x_{i+1},\ldots,x_j)$, with $j-i>1$, where $x_i, x_j \in \mathcal{N}$
 and the intermediate hidden variables $(x_{i+1},\ldots,x_{j-1})$ take values from the abnormal set. In other words, an excursion is the sub-path having  the start and end states clamped to normal states and with all intermediate variables clamped to abnormal values. Further, a special case of an excursion is a {\em restricted excursion} where the intermediate sub-path $(x_{i+1},\ldots,x_{j-1})$ is clamped to the same abnormal state.

To count excursions, we introduce a new sequence of auxiliary variables $\bfe = (e_1,\ldots,e_N)$ which aim to signify the different phases of the excursion cycle. These variables unfold sequentially given the path $\bfx$ according to the following deterministic chain. Initially, $e_1$ is set to zero so that $p(e_1|x_1) =  \delta_{e_1,0}$ and  then any subsequent $e_n$ is drawn according to  
\begin{equation}
	p(e_n| e_{n-1}, x_{n-1}, x_n) = 
	\left\{ 
		\begin{array}{cl}
			\delta_{e_n,1}          & x_{n-1} \in \mathcal{N}  \ \& \ x_{n} \in \overline{\mathcal{N}},  \\   
			\delta_{e_n,0}          & x_{n-1} \in \overline{\mathcal{N}}  \ \& \ x_{n} \in \mathcal{N},  \\      
			\delta_{e_n,e_{n-1}}    &  \text{otherwise}. \\      
		\end{array} 
	\right.
	\label{eq:excursion}
\end{equation}
Here, the first part of the conditional signals the initiation of an excursion where $e_n$ is set to one once a transition from a normal state to an abnormal state occurs. The second part signifies the end of the excursion where we return to a normal state. The third part replicates the previous value and deals simultaneously with both intermediate variables in the excursion sub-path, in  which case $e_n=e_{n-1}=1$, and situations where $\bfx$ has started in an abnormal state and an initiation of an excursion has not occurred so far, in which case $e_n = e_{n-1}=0$. The key now to count excursions is to increment a counter any time there is transition from one to zero in the path $\bfe$ signifying the completion of an excursion. This is achieved using counting variables $\bfs$ generated given $\bfe$, so that $s_1=0$ and any subsequent $s_n$ is drawn from 
\begin{equation}
	p(s_n|s_{n-1}, e_n, e_{n-1}) = I(e_{n-1}=1 \ \& \ e_n=0) \delta_{s_n,s_{n-1}+1} + \left(1 - I(e_{n-1} = 1 \ \& \ e_{n}=0) \right) \delta_{s_n,s_{n-1}}.
	\label{eq:countExcurs}
\end{equation} 
The initial HMM is augmented hierarchically with the above two layers of auxiliary variables so that  
\begin{equation}
	p(\bfy,\bfx,\bfe,\bfs) = p(\bfy|\bfx) p(\bfx) p(\bfe|\bfx) p(\bfs|\bfe), 
\end{equation}
is the joint density of the extended state-space HMM and each triple $(x_n,e_n,s_n)$ consists of the new extended hidden state. Then, by working analogously to section \ref{sec:dynprog} we can derive recursions for all types of $k$-segment inference problems associated with counting excursions. For instance, by specifying a maximum number of $k_{max}$ excursions we can introduce evidence into the final state $s_N$, such that $s_N \leq k_{max}$, and then obtain all optimal $k_{max}$ paths containing $k=1$ up to $k=k_{max}$ excursions using the Viterbi algorithm. Since each variable $e_n$ takes two possible values and $s_n$ takes $k_{max}+1$ possible values, the complexity of all dynamic  programming algorithms will be $O(2 (k_{max} + 1)M^2 N)$ which is twice as slow as generalized $k$-segment inference. 
 
Dealing with restricted excursions requires only a modification of the third  ``otherwise'' part in  eq.\ (\ref{eq:excursion}). In particular, this part must now be modified so that once an excursion cycle has previously been initiated, i.e.\ $e_{n-1}=1$, we will count any transition happening between abnormal states. More precisely, this part becomes 
\begin{equation}
	p(e_n| e_{n-1}, x_{n-1}, x_n) = I(e_{n-1} = 1 \ \& \ x_{n-1} \neq x_n) \delta_{e_n,e_{n-1}+1} + \left(1 - I(e_{n-1} = 1 \ \& \ x_{n-1} \neq x_n) \right) \delta_{e_n,e_{n-1}}.
\end{equation}
Then, the problem of counting restricted excursions is solved by constraining all $e_n$ variables to take only the two values $\{0,1\}$, so that once an excursion cycle is been initiated we cannot transit to a different abnormal state. The time complexity of the dynamic programming recursions remains $O(2 (k_{max} + 1)M^2 N)$ as in the simple excursion case. 

To illustrate the concept of extracting excursions we return to the dataset of Figure \ref{fig:illustrativePlots}, where  
we would like to count excursions so that the first and second states comprise the null set and the remaining 
third state is taken as abnormal. The panel in the last row of Figure \ref{fig:illustrativePlots} shows several optimal paths found by counting excursions 
where, for clarity, only the excursion segments are displayed using black solid lines.

\section{Relation to other methods  \label{sec:relatedwork}}

Our method formalizes and generalizes the approach of \cite{kohlmorgen2003optimal} who provided the first solution (as far as we are aware) for a specific form of the $k$-segment inference problem. \cite{kohlmorgen2003optimal} recognized that an exact dynamic programming solution for the optimal decoding MAP estimation problem existed. In this article, we have placed that insightful observation by \cite{kohlmorgen2003optimal} within a novel counting Markov chain framework and showed that the use of dynamic programming can also be used for marginalization and sampling of random variables and thus, for instance, allow the computation of marginal probabilities over subset of hidden paths using the forward recursion and simulating samples with exactly $k$ segments using the FF-BS algorithm. The use of augmentation with auxiliary variables means that our framework is easily generalizable as someone can tackle different types of inference problems by constructing suitable counting chains. For instance, in Section \ref{sec:countingsubset}, we took this forward by introducing and solving generalized $k$-segment inference problems in HMMs simply by generalizing the structure of the counting chain. 

In addition, there are similarities in the way we construct counting chains  with that of explicit duration HMMs \citep{MitchellHJ95,murphy_02,Yu10}, which consists of a modification of the original HMM where each hidden state  emits not a single observation but a sequence of observations. The number of these observations is chosen randomly from a distribution. This can be thought of as introducing duration or segment length constraints in the original HMM, so that the resulting model is a hidden semi-Markov model. From technical point the use of counting variables in ED-HMMs shares similarities with our methodology, however, the scope of our approach is very different. Specifically, in our case the counting variables are used to obtain probabilities and hidden paths in the original standard HMM, i.e.\ we do no alter the original HMM but instead we do exploratory inference in this model, while in the ED-HMM the counting variables define a new model (marginalizing out the $s_n$s from the joint prior distribution of $x_n$s and $s_n$s gives a new semi-Markov model). 

The task of $k$-segment inference in HMMs is strongly related to change point estimation. Traditional change point estimation algorithms; see e.g. \citep{AugerLawrence1989,Fearnhead2006,Fearnhead2007}, allow the computation of optimal segmentations of sequential data having one up to $k_{max}$ segments in $O(k_{max} N^2)$ time, i.e.\ these algorithms have quadratic complexity in the length of the data sequence. In contrast, our algorithms have linear complexity in the length of the sequence and therefore they can be applied to massive  datasets as those encountered in bioinformatics. Recently, \cite{KillickFE11arxiv} presented a linear complexity algorithm for change point estimation, which however, does not solve the optimal decoding problem  in $k$-segment inference (i.e.\ it does not find all optimal segmentations having one up to a maximum number of segments), but instead it discovers a single segmentation with an {\it a priori} unknown number of segments. 

\cite{YauHolmes2013} also developed a decision theoretic approach for segmentation using Hidden Markov models by defining a loss function on transitions and identifying a Viterbi-like dynamic programming algorithm to efficiently compute the hidden state sequence that minimizes the posterior expected loss. The properties of the sequence predictions are modified through specification of the loss penalties on transitions as supposed to altering the transition dynamics of the Hidden Markov model. The $k$-segment algorithms developed here can also be applied to the method of \cite{YauHolmes2013} to produce sequence predictions that minimize the posterior expected loss criterion subject to a desired $k$-segments constraint. The combination of the decision theoretic approach and the use of $k$-segments therefore provides a powerful tool for exploration of the complete state space of Hidden Markov models.

\section{Examples \label{sec:experiments}} 

Next, we demonstrate the utility of $k$-segment methods in two real-word applications. Specifically, in section \ref{sec:genomics} we consider the problem of copy number identification
in cancer genomic sequences, while in section \ref{sec:topics} we discuss an application to text retrieval and topic modelling.

\subsection{Genome-wide DNA copy number profiling in cancer \label{sec:genomics} }

In this section, we consider the problem of genome-wide classification of somatic DNA copy number alterations (SCNAs) in cancer. SCNAs are a important constituent of the mutational landscape in cancer and refer to numerical copy number changes that result in extra or lost copies of parts of the genome. In cancer, these alterations lead to the loss of tumor suppressor genes (which restrict tumorigenic activity) or the gain of oncogenes (which promote tumorigenic activity) and many copy number alterations of such genes have been identified as being associated with cancer \citep{Beroukhim2010a}. Next generation sequencing or microarray technologies have allowed cancers to be probed on a genome-wide scale for SCNAs and a number of statistical models have been developed to support the analysis of this data \citep{Loo2010,Yau2010,Chen2011,Carter2012,Yau2013}. A particularly popular class of these models have utilised Hidden Markov models to model microarray intensities or sequencing reads as observations of a hidden (discrete) state process that corresponds to the unobserved copy number sequence. 

Specifically, a single nucleotide polymorphism (SNP) microarray dataset consists of a sequence of bivariate measurements $\{\bm{y}_i \}_{i=1}^n$ at $n$ SNP locations spread across the genome. The first dimension of the measurements known sometimes as the {\it Log R Ratio} values which are intensity measurements whose magnitude is proportional to the total copy number at that particular genomic location. In human genome analysis, the Log R Ratio values are typically normalized such that values approximately equal to zero correspond to a DNA copy number of two since we typically inherit one copy of every gene from each parent. The second dimension, sometimes known as the {\it B allele frequency}, measures the relative contribution of one of the parental alleles to the overall signal which can allow us to determine which parental allele is lost or gained. 

In \cite{Yau2010}, these data sequences are modelled using a Bayesian hierarchical model specified via the following relationships:
\begin{align}
	\bm{y}_i | x_i, \bm{m}, \Sigma, \nu & \sim \mathrm{Student}( \bm{m}_{x_i}, \Sigma_{x_i}, \nu ), ~ i = 1, \dots, n, \\
	x_i | x_{i-1} & \sim \mathrm{Multinomial}( \bm{\pi}_{x_{i-1}} ),
\end{align}
where $x_i \in \{ 1, \dots, S \}$ denotes the copy number state at the $i$-th location, $\{ \bm{m}_j, \Sigma_j \}$ denotes the expected signal measurements and noise covariance for the $j$-th copy number state and $\bm{\pi}$ is a transition matrix such that $\bm{\pi}_{j}$ corresponds to the transition probabilities out of the $j$-th copy number state. Note, we present only an abbreviated and simplified version of the complete model by \cite{Yau2010} here. For full details, see the original reference. 

Table \ref{tab:copynumberstates} shows an example set of copy number states. \cite{Yau2010} models transitions between super-states as relatively unlikely events leading to a ``sticky" HMM that produces relatively few super-state segments. Dynamics within super-states are modelled via an embedded Markov chain that approximates the patterns of genotypes observed in real data. The primary scientific interest is in the switching between super-states but it is necessary to fully model the complete genotypes in order to achieve this.

\begin{table}[!h]
	\centering
	\begin{tabular}{|c|c|c|c|c|}
	\hline 
	Copy Number State & Total Copy Number & LOH & Genotype & Super-state \\ 
	\hline 
	1 & 0 & N/A & N/A & 1 \\ 
	\hline 
	2 & 1 & 0 & A & 2 \\ 
	\hline 
	3 & 1 & 0 & B & 2 \\ 
	\hline 
	4 & 2 & 0 & AA & 3 \\ 
	\hline 
	5 & 2 & 0 & AB & 3 \\ 
	\hline 
	6 & 2 & 0 & BB & 3 \\ 
	\hline 
	7 & 3 & 0 & AAA & 4 \\ 
	\hline 
	8 & 3 & 0 & AAB & 4 \\ 
	\hline 
	9 & 3 & 0 & ABB & 4 \\ 
	\hline 
	10 & 3 & 0 & BBB & 4 \\ 
	\hline 
	11 & 2 &1 & AA & 5 \\ 
	\hline 
	12 & 2 & 1 & BB & 5 \\ 
	\hline 
	\end{tabular} 
	\caption{Example copy number states. Each copy number state is associated with a total copy number and genotype which tells us the number of each parental allele (A/B). The super-state corresponds to subsets of copy number states with identical total copy number and/or loss of heterozygosity (LOH) status. }
	\label{tab:copynumberstates}
\end{table}

\begin{figure}[!h]
\centering
\begin{tabular}{c}
{\includegraphics[width=\textwidth]{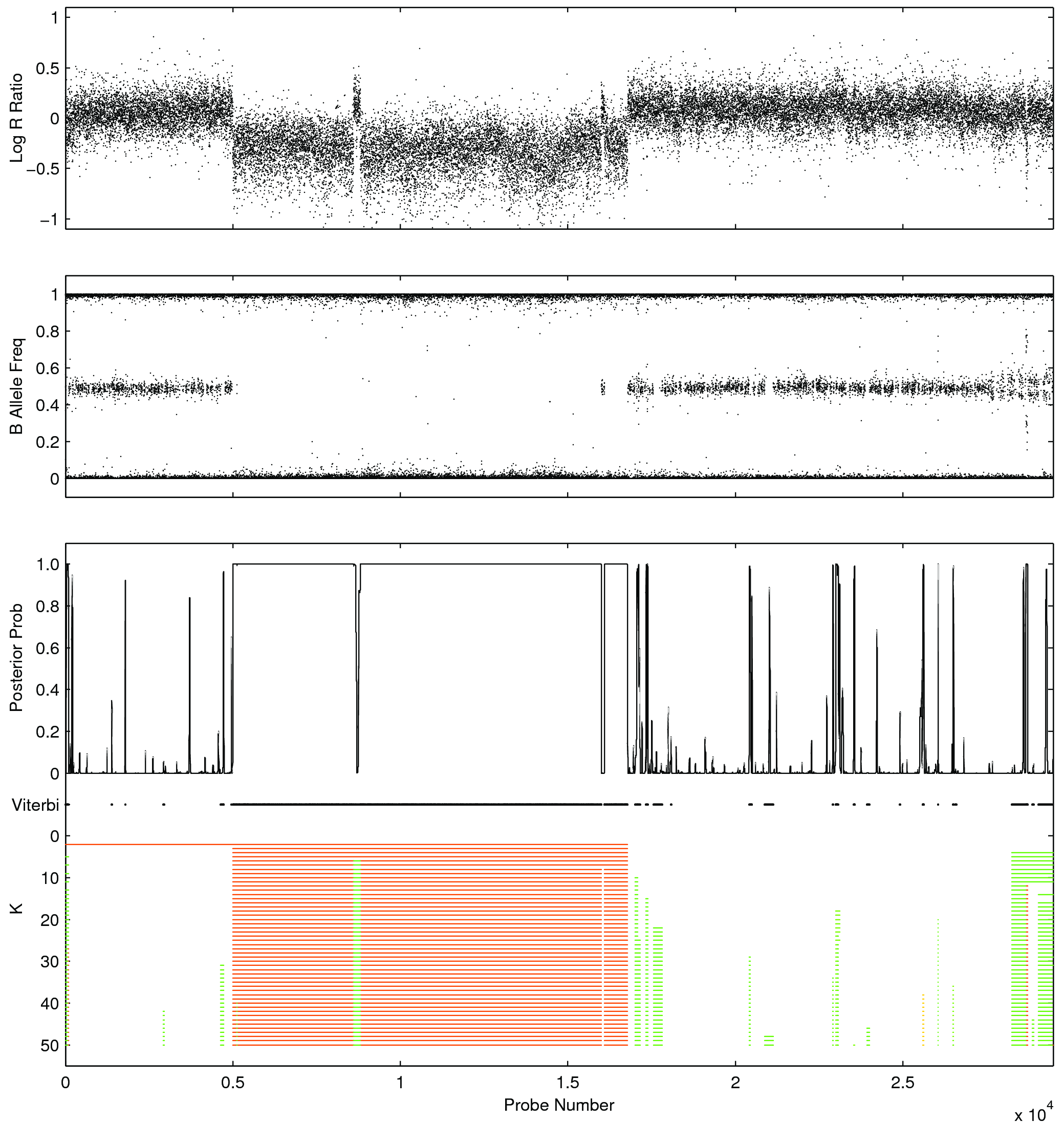}}
\end{tabular}
\caption{Copy number analysis of the colorectal cancer cell line SW837 (Chromosome 1) using site-wise marginal posterior probabilities of a copy number aberration from the Forward-Backward algorithm, the Viterbi algorithm (black lines indicate detected regions of aberrant copy number), and $k$-segment analysis for different fixed super-state segment numbers.}
\label{fig:kseg-genomics-ex}
\end{figure}

Full Bayesian posterior inference for this type of model is prohibited by the size of the datasets ($O(n) \approx 10^6$). \cite{Yau2010} perform model fitting using expectation-maximization to compute MAP parameter estimates and condition on these to obtain MAP segmentations using the Viterbi algorithm. The forward-backward algorithm can also be applied to obtain site-wise posterior probabilities of state occupation. Figure \ref{fig:kseg-genomics-ex} shows an example copy number analysis of chromosome 1 of a colorectal cancer cell line SW837 from a SNP microarray dataset using the OncoSNP software from \cite{Yau2010}. The chromosome exhibits a number of copy number alterations leading to changes in the pattern of the Log R Ratio and B Allele Frequency along the chromosome. Genomic regions with non-normal total copy number (2) can be identified from the Viterbi segmentations and the site-wise posterior probabilities. 

The application of our $k$-segments methods can be used to augment these standard analyses with additional exploratory information. Figure \ref{fig:kseg-genomics-ex} shows segmentations conditional on different fixed super state segment numbers obtained using $k$-segments. Here, we have used the ability to count certain transitions in $k$-segment inference to good effect to count only transitions between super-states and exclude transitions between copy number states within super-states. This means the $k$-th segmentation represents the most probable copy number segmentation that involves $k$ different super-state segments as supposed to $k$ segments defined on the original state space which would include transitions between states within super-states. These segmentations allow the exploration of alternative segmentations that differ from the MAP solution and yet retain segmental constraints that cannot be observed from the site-wise marginal probabilities.

\subsection{Application to text retrieval using hidden Markov topic models  \label{sec:topics}}

In this section, we apply $k$-segment inference to a information 
retrieval task where the objective is to process long documents and 
extract segments referring to certain topics. For this purpose, we define    
a hidden Markov topic model, as those proposed in \citep{Gruber2007, Andrews2010}, 
that builds upon popular topic models such as probabilistic 
latent semantic indexing \citep{Hofmann2001} and latent Dirichlet allocation \citep{Blei2003}. 
These latter models 
represent a document as a set of words $\bfy_d = (y_{d,1}, \dots,y_{d,N_d})$ 
where each $y_{d,n} \in \{1,\ldots,V\}$ points into
a vocabulary consisting of $V$ keywords. The words  are generated 
exchangeably from a mixture distribution where mixing proportions are 
document-specific parameters while mixture components are multinomial distributions over the 
vocabulary. The latter distributions aim at capturing  semantic 
topics, such as {\em politics}, {\em sports}, {\em food} and etc. 
Learning in these models is typically carried out in an unsupervised manner by using a 
corpus of several documents and assuming a common pool of topics so that each document, through 
the document-specific mixing proportions, can contain only a subset of topics. 
The basic topic models ignore word ordering and do not model spatial correlation 
according to which nearby words are more likely to belong to the same topic. Motivated by this limitation, 
HMM-based extensions have been developed in \citep{Gruber2007, Andrews2010} that assume that the latent topics 
of words in ordered text follows a Markov chain. Next, we construct a similar HMM 
and use it to solve an information retrieval task based on semi-supervised learning. 

Assume an unknown-content (test) document $d$ in which we would like to scan through and retrieve 
segments referring to certain topics. As before the document is represented by a set of words
$\bfy_d = (y_{d,1}, \dots,y_{d,N_d})$ which are ordered according to their 
appearance in the text and  assumed to have been generated from an HMM. Specifically,  we assume there is a 
path $\bfx_d = (x_{d,1}, \dots,x_{d,N_d})$ such that each  
$x_{d,n} \in \{1,\ldots,M \}$ indicates the hidden topic of word $y_{d,n}$. Further, the set of these topics is divided into the {\em relevant topics} and the {\em irrelevant topics} with the relevant topics being the ones from which 
we wish to extract text segments, estimate posterior probabilities of appearance and etc, while 
the irrelevant topics are unknown and document-specific topics of no interest to us. Without loss of generality, and to 
simplify our presentation, we shall assume $M=2$ so that there is a one relevant 
and one irrelevant topic. The relevant topic is described by multinomial parameters  
$\bfphi_{r} = (\phi_{r,1}, \ldots, \phi_{r,V})$ so that the emission distribution that 
generates a word $y_{d,n}$ is such that
\begin{equation}
p(y_{d,n} | x_{d,n} = 1) = \phi_{r,y_{d,n}}.
\end{equation}
$\bfphi_r$ is assumed to have been estimated by supervised learning 
using fully labeled documents according to the equations: 
\begin{equation}
\phi_{r,v} = \frac{n_{v} + 1}{n + V}, \ \ v=1,\ldots,V,
\label{eq:reltopic}
\end{equation}
where $n_{v}$ is the number of times the $v$th word appears in the labeled data and $n$ is the total number 
of words in these data. Notice that the above is simply the Bayesian mean estimate under an uniform 
Dirichlet prior over $\bfphi_r$. Similarly, the emission distribution for the irrelevant topic, i.e.\ 
$p(y_{d,n} | x_{d,n} = 2)$, is described by the parameter vector
$\bfphi_d = (\phi_{d,1}, \ldots, \phi_{d,V})$ which is a document-specific parameter to be estimated. 
Furthermore, the prior distribution $\bfpi_d$  
and transition matrix $A_d$ of the HMM are also document-specific parameters and the full set $(\bfphi_d, \bfpi_d, A_d)$ can be 
estimated via the EM algorithm while $\bfphi_r$ is kept fixed. 
In practice, we also place a conjugate Dirichlet prior over all unknown parameters so 
that EM finds MAP point estimates similar to those of eq.\ (\ref{eq:reltopic}).

In the remaining of this section we demonstrate the above system using a freely available text corpus taken from the University 
of Oxford electronic library.\footnote{See {\tt http://www.bodleian.ox.ac.uk/ora}.} Specifically,    
we collected a set of $119$ doctoral theses on several subjects such as History, Social Sciences, Philosophy, Law, Politics, Literature
and Economics. The topic of Economics was considered to be the relevant topic while all remaining topics were taken 
as irrelevant. Ten out of $119$ documents were classified (according to the library database system) to be about 
Economics while the remaining $109$ theses were scattered across the other topics. Each $d$th document was represented by a sequence 
of words from a dictionary of size $V=1260$  which was defined separately by choosing all different words 
from a large set of freely accessible Wikipedia articles.\footnote{Following also the standard practise in topic modelling to exclude from 
the vocabulary very common words, of non semantic meaning, 
such as 'the', 'of', 'and' etc.} The multinomial parameters for the relevant topic of Economics was obtained by
supervised learning using counts of words obtained from a small set of Wikipedia entries such as the entries Economics, Finance and Investment. 
Having preprocessed each document as above, we then considered two types of prediction tasks: i) classification and ii) detection that we describe next in turn.    

\begin{figure}[!htb]
\centering
\begin{tabular}{c}
{\includegraphics[scale=0.5]
{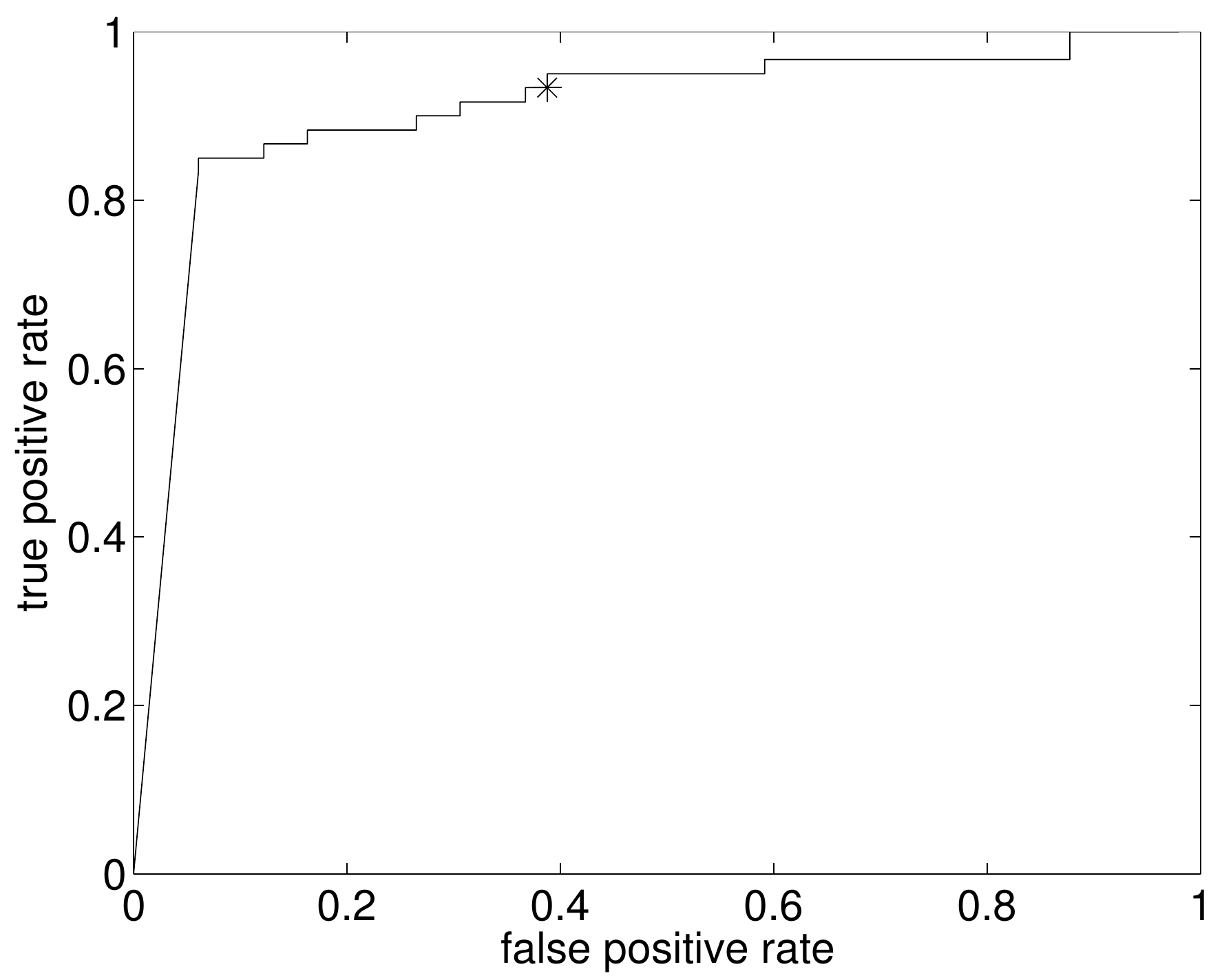}} 
\end{tabular}
\caption{The blue line shows the ROC (receiver operating characteristic) curve for the $k$-segment method that classifies  
documents by using the topic-occurrence posterior probability $p(c_{\bfx}>0|\bfy)$. The black star shows the pair of the true and false positive rates 
corresponding to the classifier obtained by the standard Viterbi MAP path.}
\label{fig:topics1}
\end{figure}

{\bf Classification.} For the classification task the objective was to predict in a test document the presence or absence of at least one
occurrence of a segment from the topic Economics. The test documents consisted of the $109$ theses, originally annotated as non-Economics documents,
that were randomly perturbed in order to create a ground-truth dataset of known classification as explained in the Appendix  \ref{app:text}.
Given this test dataset, the objective was to construct a binary classification system and classify each of the documents as 
relevant, i.e.\ as containing at least one text segment about Economics, or as irrelevant. Each test document was processed separately 
by applying the EM algorithm discussed earlier. Then, to achieve probabilistic classification, 
the posterior probability for the occurrence of at least one segment from the relevant topic is required. It can be obtained 
by applying k-segment inference using a counting variable $c_{\bfx}$ that 
increments only when a segment from the relevant topic occurs. Notice that this requires the use of generalized counting, as described in section 
\ref{sec:countingsubset}, that uses certain values for the constraints $\bfmu$ and $C$.\footnote{Assuming that the first hidden state in the HMM corresponds to the 
relevant topic and the second one to the irrelevant topic, $\bfmu = [1 \ 0]$ while the first row of $C$ is $[0 \ 0]$ and the second row 
$[1 \ 0]$.} Then, the posterior probability $p(c_{\bfx}>0|\bfy_d)$ 
is computed using the forward pass in the augmented HMM which subsequently provides a probabilistic classifier. Using different thresholds 
in the classification probability, we can obtain different decision systems of varying false positive and true positive rates as shown by the ROC curve 
in Figure \ref{fig:topics1}. In contrast, if we were about to perform classification using the Viterbi MAP path 
we can only obtain a single decision system that classifies documents as relevant or irrelevant based upon whether a 
segment from the relevant topic occurs or not in the Viterbi path. Such system gives a single value for the true positive 
and false positive rate as shown in Figure \ref{fig:topics1}. Clearly, $k$-segment's 
ability to compute non-trivial posterior probabilities allows for more flexible uses of HMMs when building decision making systems.  

{\bf Detection.} We now turn into the second task which is concerned with the detection 
of individual segments within a document that belong to the relevant topic. We adopt a standard information retrieval setup 
that is referred to as top-k retrieval \citep{InformationRetrieval2010}. This is the task  of retrieving k patterns (typically full documents)
that are most relevant to a given query among a large set of other possible patterns. Our specific top-k retrieval task will be to extract 
top-k text segments within the same large document and to achieve that we shall use the hidden Markov topic model. 
Also, to account for documents that may contain fewer than $k$ segments from the relevant topic, we will relax the constraint 
to retrieve exactly $k$ segments to the softer constraint of retrieving at most $k$ segments. 
It is worth noticing that there is a similarity of $k$-segment 
problems in HMMs and top-$k$ retrieval since both involve inference under counting constraints. More precisely, 
$k$-segment methods can naturally tackle the previous top-$k$ retrieval task by applying optimal decoding, under the constraint 
$c_{\bfx} \leq k$, that finds the optimal hidden path containing at most $k$ text segments associated 
with the relevant topic. Next, it order to evaluate such system in test documents with known ground-truth segments,  
we randomly perturbed the $109$ test documents as explained in the Appendix \ref{app:text}. 

\begin{figure}[!htb]
\centering
\begin{tabular}{cc}
{\includegraphics[scale=0.4]
{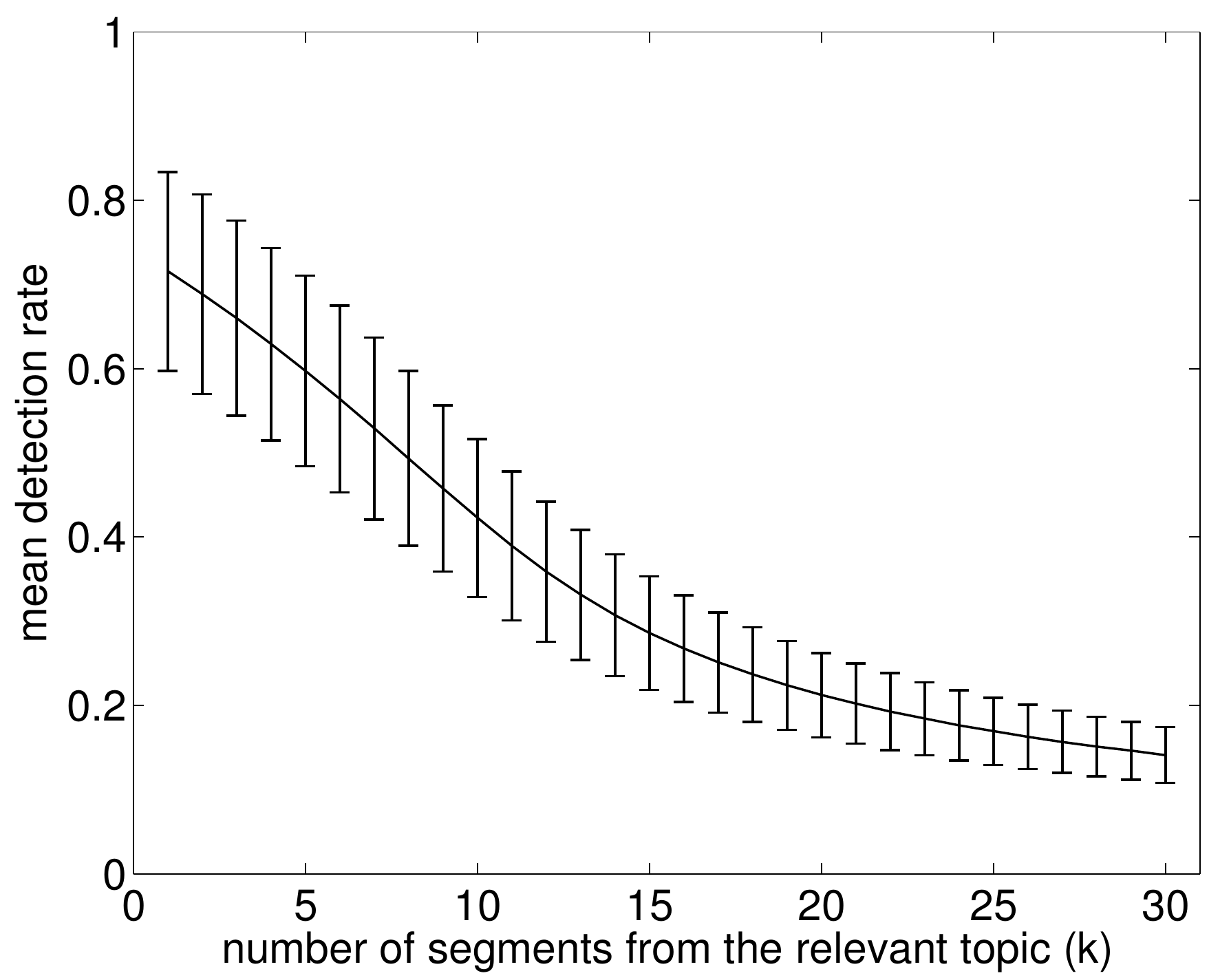}} & 
{\includegraphics[scale=0.4]
{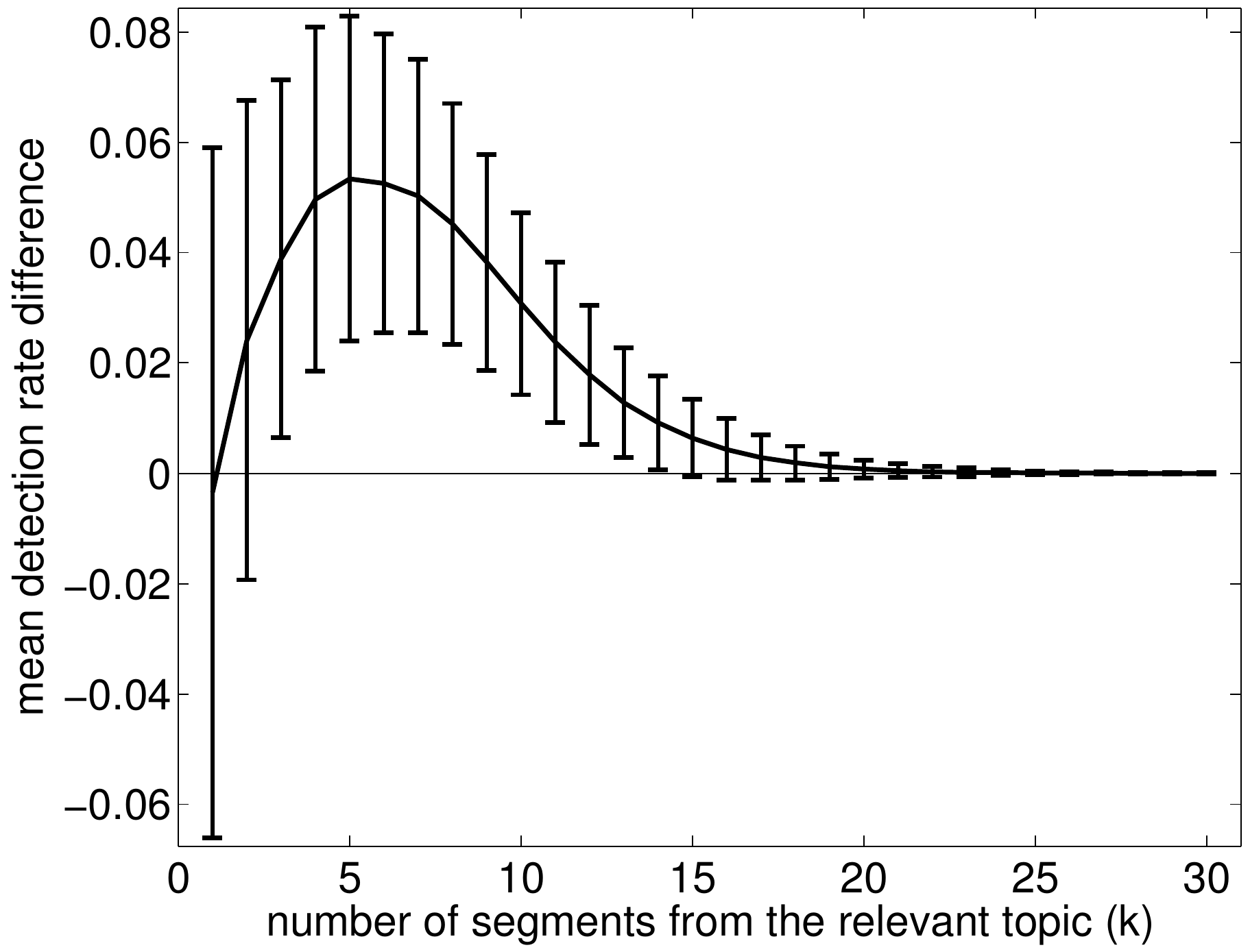}} \\
(a)  & (b)
\end{tabular}
\caption{The panel in (a) shows mean detection rates (solid central line) for top-$k$ systems, obtained by varying $k$, 
together with $95\%$ confidence intervals computed by repeating the experiment $100$ times.
The panel in (b) shows mean differences in detection rates of the $k$-segment method and standard Viterbi together with $95\%$ 
confidence intervals obtained by the $100$ repeats. 
}
\label{fig:topics2}
\end{figure}

To measure performance, we make use of a popular evaluation measure used in visual object detection literature. 
More precisely, detecting segments of certain topics in documents is similar to detecting instances of object categories in natural images. 
There, the detection problem is to predict a bounding box that locates an instance of an object category within the image. The  well-established evaluation measure, 
used in the PASCAL visual object recognition challenge \citep{Everingham:2010}, is the overlap area ratio.  
Adopting this in our case, we have that for a predicted segment $S_p = [i_l, i_r]$, 
where $i_l$ and $i_r$ are the segment start and end locations within the test document, the overlap ratio is defined by
\begin{equation}
r = \frac{|S_p \cap S_{gt}|}{|S_p \cup S_{gt}|}. 
\end{equation}
Here, $S_{gt}$ is the ground-truth segment, $S_p \cap S_{gt}$ is the intersection of the predicted and the ground segments  
and $S_p \cup S_{gt}$ is their union. Clearly, $r \in [0,1]$ and values close to zero indicate poor detection while values close to one
indicate strong detection. We consider as correct detections all cases when $r$ exceeds the threshold of $80\%$; 
for an illustrative example of a correct detection see Figure \ref{fig:textdetectionExample}.
Also, to get a total document-specific performance that is normalized with respect to $k$, we average according to  
\begin{equation}
\text{per document detection rate} = \frac{1}{k} \sum_{i=1}^{k_p} I(r_i > 0.8),
\end{equation}
where $k_p \leq k$ is the number of predicted segments. From this we can obtain a mean detection rate that gives 
the overall performance in the whole test dataset. Figure \ref{fig:topics2}(a) shows means detection rates for several top-$k$ systems
of varying values of $k$. Confidence intervals were obtained by repeating the experiment $100$ times, so that in each repeat 
a random test dataset of $109$ documents was created using bootstrapping 
together with the standard randomization involved in the segment insertion (see Appendix \ref{app:text}). 

Furthermore, it is interesting to compare the $k$-segment based method with a system constructed using the standard Viterbi MAP path in 
the HMM. Standard Viterbi gives a single path that will contain a priori an unknown number
of segments from the relevant topic. Thus, to get top-$k$ retrieval systems (for different values of $k$), we can rank all relevant-topic
segments with respect to their length so that the top-1 retrieval system simply outputs the longest segment in the list, the top-$2$ retrieval system 
outputs the two longest segments and so forth. Using the same bootstrapped $100$ repeats we also evaluated the 
standard Viterbi system and for each repeat we recorded the difference in mean detection rates ($k$-segment rate minus the standard Viterbi rate).  
Figure \ref{fig:topics2}(b) displays the mean of these differences together with $95\%$ confidence intervals and for several values of $k$.
Clearly, there is a certain range of
$k$ values where the $k$-segment method outperforms the standard Viterbi method. Moreover, as $k$ increases, the $k$-segment  
constraint $c_{\bfx} \leq k$ becomes weaker and the corresponding optimal paths converge to the standard Viterbi MAP paths which 
explains the fact that the performance of the two methods becomes identical for large $k$. 

To summarise, both tasks in text retrieval presented above indicate that $k$-segment inference allows for more flexible use of 
HMMs which provides us with new options when building classification and decision making systems.    

\begin{figure}
\begin{framed}
... changes in services had brought within the direct employ of local authorities new professional groups who could claim authority and expertise in the field of child welfare - school medical officers, school nurses, juvenile employment workers, in addition to the many volunteer roles which were integral to the operation of Education Departments at local level. These new professional groups ...
\end{framed}  
\begin{framed}
... changes in services had brought within the direct employ of \textcolor{red}{complexity system modeling to model market communication networks. This would be very useful to understand the connection between price signaling, public awareness and regulatory demand. In addition to modeling market networks,}  of Education Departments at local level. These new professional groups ...
\end{framed}  
\begin{framed}
... changes in services had brought within the direct employ of complexity system modeling  \textcolor{blue}{to model market communication networks. This would be very useful to understand the connection between price signaling, public awareness and regulatory demand. In addition to modeling market networks,}  of Education Departments at local level. These new professional groups ...
\end{framed}  
\caption{An example of detection of a text segment from the relevant topic of Economics. 
The box on top displays a piece of text from  a test document  before we randomly inserted segments from 
the topic of Economics (see Appendix \ref{app:text}). The middle box displays the same text after having randomly inserted (and replaced the original piece of text) a segment from the topic of Economics which is shown in red. The box at the bottom shows in blue color the segment predicted as belonging to the relevant topic. In this case, the predicted segment was classified as a correct detection since it overlaps more than $80\%$ with the ground-truth segment shown in the middle box.}
\label{fig:textdetectionExample}
\end{figure}

\section{Discussion  \label{sec:discussion}}

HMMs can allow for highly efficient analysis of large quantities of sequence data. However, existing methods for reporting of posterior summaries from HMMs such as the Viterbi MAP path and the marginal probabilities are rather blunt. Here, we demonstrated how the use of auxiliary counting variables allows for computationally efficient exploration of the model fit using $k$-segment algorithms. It is important to note that the techniques we developed are generic and the augmentation scheme can be applied either {\it a posteriori} to HMMs already fitted to data or {\it a priori} during model fit. In cancer genomics, $k$-segment inference can be an useful exploratory tool that can help researchers to analyze genomic sequences in different resolutions or target events of particular types, facilitating thus the process of getting novel insight into structural rearrangements in cancer genomes. For other type of applications, that appear for instance in machine learning and pattern recognition, the proposed methods can allow to build more flexible HMM-based classification and decision making systems, as we have demonstrated using the text retrieval example. 
  
Regarding future work, an interesting research direction is to exploit the ability of $k$-segment inference to efficiently explore the HMM posterior distribution in order to provide input into constructing meta statistical models. For instance, the ability to obtain alternative explanations of the same data sequence that may have high utility to the research scientist but occurs with very low probability. could allow the practitioner to re-rank different explanations based on his expertise and subsequently provide feedback into the model that can be used for supervised re-training. 
A second future direction is to exploit the fact that  
training an HMM under a $k$-segment constraint often gives sparse transition matrices.
This is not surprising, since a bound on the number of segments essentially limits 
the number of transitions along the hidden path which subsequently can result in many inferred zeros in the 
transition matrix. Such sparsity property could be useful when we would like to perform {\em state selection} 
in large state-space HMMs.    

To conclude, as data sets become larger and models more complex we expect to see increasing need for computationally efficient methods 
for posterior model exploration and statistical inference under constraints. In this paper, we have presented one such approach that significantly  expands the statistical algorithmic toolbox of HMMs.

\section{Acknowledgements \label{sec:acknow}}

MT and CH were supported by a Wellcome Trust Healthcare Innovation Challenge Fund award (Ref No. HICF-1009-026) and the Lincoln College Michael Zilkha Fund. MT was also supported from ``Research Funding at AUEB for Excellence and Extroversion, Action 1: 2012-2014''. 
CY was funded by a UK Medical Research Council Specialist Training Fellowship in Biomedical Informatics (Ref No. G0701810) and a New Investigator Research Grant (Ref No. MR/L001411/1).

\bibliographystyle{apalike}

\appendix 

\section{Proofs for the auxiliary variable reformulation of $k$-segment problems \label{app:proofs}}

Here, we provide proofs for the correctness of the reformulation of the three $k$-segment inference problems presented in section \ref{sec:counting}. 
Firstly, we will show that  $p(\bfx|\bfy, s_N=k)$, computed via the augmented HMM, 
is equal to $p(\bfx|\bfy,c_{\bfx}=k)$ given by eq.\ (\ref{eq:pxEKy}). We have that $p(\bfx|\bfy, s_N=k)$
is defined by 
\begin{equation}
p(\bfx|\bfy, s_N=k) \propto
p(\bfy|\bfx) p(\bfx) \sum_{\bfs_{\setminus N}} p(\bfs_{\setminus N}, s_N=k|\bfx).
\label{eq:pxysNK}
\end{equation}
What we  need to show is that $\sum_{\bfs_{\setminus N}} p(\bfs_{\setminus N}, s_N=k|\bfx)$ 
is equal to the indicator function $I(c_{\bfx} = k)$.  Since $p(\bfs_{\setminus N}, s_N=k|\bfx)$ is a deterministic 
distribution, given that $\bfx$ has $k$ segments 
there will be an unique $\bfs_{\setminus N}^*$ such that 
$p(\bfs_{\setminus N}^*, s_N=k|\bfx)=1$ and zero for all remaining  
$\bfs_{\setminus N}$s. If $\bfx$ does not contain $k$ segments, $p(\bfs_{\setminus N}, s_N=k|\bfx)=0$ for any 
$\bfs_{\setminus N}$.  Thus, when $\bfx$ has $k$ segments
 $\sum_{\bfs_{\setminus N}} p(\bfs_{\setminus N}, s_N=k|\bfx) = 1$, otherwise 
$\sum_{\bfs_{\setminus N}} p(\bfs_{\setminus N}, s_N=1|\bfx) = 0$. Therefore,
$\sum_{\bfs_{\setminus N}} p(\bfs_{\setminus N}, s_N=1|\bfx) = I(c_{\bfx} = k)$ for any $\bfx$, 
from which we conclude that $p(\bfx|\bfy, s_N=k)$ reduces to the definition  
of $p(\bfx|\bfy, c_{\bfx}=k)$ from eq.\ (\ref{eq:pxEKy}). From that, we can immediately obtain that the 
term that normalizes the right hand side of (\ref{eq:pxysNK}), i.e.\ the quantity $p(s_N=k,\bfy)$, 
is equal to $p(c_{\bfx}=k,\bfy)$. This completes the proof regarding the correctness of the
probability computation.

Based on the above, we can also conclude that the initial  optimal decoding solution  $\bfx^*$ is the MAP of 
$p(\bfx|\bfy, s_N=k)$, i.e.\  
\begin{equation}
\bfx^* = \arg  \max_{\bfx} \left[
p(\bfy|\bfx) p(\bfx) \sum_{\bfs_{\setminus N}} p(\bfs_{\setminus N}, s_N=k|\bfx)\right].
\end{equation}
Given now that $p(\bfs|\bfx)$ is a deterministic distribution the 
sum operation can be replaced by a max operation so that 
\begin{equation}
\bfx^* = \arg  \max_{\bfx} \left[
p(\bfy|\bfx) p(\bfx) \max_{\bfs_{\setminus N}} p(\bfs_{\setminus N}, s_N=k|\bfx)\right],
\end{equation} 
or 
\begin{equation}
(\bfx^*, \bfs^*_{\setminus N}) = \arg  \max_{\bfx, \bfs_{\setminus N}} \left[
p(\bfy|\bfx) p(\bfx) p(\bfs_{\setminus N}, s_N=k|\bfx)\right],
\label{eq:maxgivensnK}
\end{equation}
which shows that the reformulated optimal decoding problem is equivalent to the initial one.

Finally, regarding path sampling, the FF-BS in the augmented HMM 
gives  a pair of paths $(\widetilde{\bfx}, \widetilde{\bfs}_{\setminus N})$ that jointly comprise an   
independent sample from $p(\bfx,\bfs_{\setminus N}|s_N=k,\bfy)$. Thus, 
$\widetilde{\bfx}$ alone is an independent sample from $p(\bfx|s_N=k,\bfy)$.

\section{Simulation of ground-truth datasets for the text retrieval example \label{app:text}}

For the classification task we created the ground-truth dataset as follows. 
For each test document sequence $\bfy_d$ we decided with probability $0.5$ 
to insert a number of $1 + g_d$ (with $g_d \sim \text{Pois}(2)$) segments from the subject Economics so that these segments had random lengths from $[10, 200]$ and were also placed in random locations within the sequence $\bfy_d$, replacing thus the original text and with the only constraint that they didn't overlap with each other. Each such set of $1 + g_d$ artificially inserted segments were also randomly selected from the $10$ theses in Economics by first picking  a thesis and then selecting $1 + g_d$ non-overlapping segments within that thesis text sequence. The whole procedure created a new dataset of $109$ documents so that a subset of them contained segments from the relevant topic and the remaining ones did not. 

For the detection task we worked similarly with the classification task discussed earlier. 
Particularly, again we randomly perturb the $109$ test documents and insert a number of $g_d \sim \text{Pois}(10)$ 
segments from the subject Economics in each of the them. The insertion of segments was done exactly as described above 
with the only difference being that now we insert segments in all documents and their number can be much larger 
since $g_d \sim \text{Pois}(10)$.

\end{document}